\documentclass[12pt]{article}
\usepackage[table]{xcolor}
\usepackage{amsmath,amssymb,exscale}
\usepackage{graphicx}
\usepackage{epsfig}
\usepackage{multicol}
\usepackage{color}
\usepackage{mathrsfs}
\usepackage{blindtext}
 \usepackage{fancyhdr}
\usepackage{hyperref}
\usepackage{cite}
\usepackage{mathtools}
\usepackage{amsmath}
\usepackage{rotating,slashed,amsmath,charter,xcolor,catchfilebetweentags,ifluatex}

\usepackage{graphicx}
\usepackage{sidecap}

\usepackage[latin1]{inputenc} 
\usepackage{multicol}  
 
 \usepackage{titlesec}
 
 \usepackage{rotating,slashed,xcolor,amsfonts,expdlist,charter}

\numberwithin{equation}{section}

\usepackage{xcolor}
\usepackage{sectsty}

\usepackage{mdframed}
\usepackage{titletoc}

\definecolor{secnum}{RGB}{13,151,225}
\definecolor{ptcbackground}{RGB}{212,237,252}
\definecolor{ptctitle}{RGB}{0,177,235}

\titlecontents{lsection}
  [5.8em]{\sffamily}
  {\color{secnum}\contentslabel{2.3em}\normalcolor}{}
  {\titlerule*[1000pc]{.}\contentspage\\\hspace*{-5.8em}\vspace*{5pt}%
    \color{white}\rule{\dimexpr\textwidth-15.5pt\relax}{1pt}}

\usepackage{hyperref}
\hypersetup{colorlinks,bookmarksopen,bookmarksnumbered,citecolor=rossos,
linkcolor=redy,pdfstartview=FitH,urlcolor=green-go}
\usepackage{slashed}

\definecolor{blus}{cmyk}{1,0.9,0,0.1}
\definecolor{verdes}{cmyk}{0.99,0,0.59,0.65}
\definecolor{rossos}{cmyk}{0,1,1,0.55}
\definecolor{redy}{cmyk}{0,1,1,0.7}
\definecolor{greeny}{cmyk}{0.99,0,0.59,0.98}
\definecolor{green-go}{cmyk}{0.79,0,0.59,0.5}

\usepackage{titlesec}

\def\Lag{\mathscr{L}}

\newcommand{\beq}{\begin{equation}}
\newcommand{\eeq}{\end{equation}}

\def\hhref#1{\href{http://arxiv.org/abs/#1}{arXiv:#1}} 

 \def\Lag{\mathscr{L}}
 
\newcommand{\tmtextbf}[1]{{\bfseries{#1}}}
\newcommand{\tmtextrm}[1]{{\rmfamily{#1}}}
\newcommand{\bp}{\bar M_P}

\def\be{\begin{equation}}
\def\ee{\end{equation}}
\def\ba{\begin{array} }
\newcommand{\Tr}{\,{\rm Tr}}
\def\bac{\begin{array} {c}}
\def\bacc{\begin{array} {cc}}
\def\baccc{\begin{array} {ccc}}
\def\bacccc{\begin{array} {cccc}}
\def\ea{\end{array}}
\def\bea{\begin{eqnarray}}
\def\eea{\end{eqnarray}}

\definecolor{red}{rgb}{1,0,0}

\def\psl{\hbox{\hbox{${p}$}}\kern-1.9mm{\hbox{${/}$}}}
\def\dsl{\hbox{\hbox{${\partial}$}}\kern-2.2mm{\hbox{${/}$}}}
\def\Dsl{\hbox{\hbox{${D}$}}\kern-2.6mm{\hbox{${/}$}}}

\def\Lag{\mathscr{L}}

\newcommand{\gappeq}{{\rlap{{\raise}.5ex\text{\ensuremath{>}}}{{\lower}.5ex\text{\ensuremath{\sim}}}}}
\newcommand{\lappeq}{{\rlap{{\raise}.5ex\text{\ensuremath{<}}}{{\lower}.5ex\text{\ensuremath{\sim}}}}}
\newcommand{\I}{\tmtextrm{1{\kern}-.24em l}}

\begin{document}
\topmargin -1.0cm
\oddsidemargin 0.9cm
\evensidemargin -0.5cm

{\vspace{-1cm}}
\begin{center}

\vspace{-1cm}

 {\Huge \tmtextbf{ 
\color{blus} Quadratic Gravity}} {\vspace{.5cm}}\\
 
\vspace{1.9cm}

{\large  {\bf Alberto Salvio }
{\em  

\vspace{.4cm}
CERN, Theoretical Physics Department, Geneva, Switzerland\\ 

\vspace{0.4cm}

\small{Preprint number: CERN-TH-2018-099}

\vspace{0.2cm}

 \vspace{0.5cm}
}

\vspace{.3cm}

\vspace{0.5cm}

}
\vspace{0.cm}

%
%
%
%
%

\end{center}

%
%
\noindent --------------------------------------------------------------------------------------------------------------------------------

\begin{center}
{\bf \large Abstract}
\end{center}
Adding terms quadratic in the curvature to the Einstein-Hilbert action renders gravity renormalizable. This property is preserved in the presence of the most general renormalizable couplings with (and of) a generic quantum field theory (QFT). The price to pay is a massive ghost, which is due to the higher derivatives that the terms quadratic in the curvature imply. In this paper the quadratic gravity scenario is reviewed including recent progress on the related stability problem   of higher derivative theories.  The renormalization of the theory is also reviewed and the final form of the full renormalization group equations  in the presence of a generic renormalizable QFT is presented. The theory can be extrapolated up to infinite energy through the renormalization group if all matter couplings flow to a fixed point (either trivial or interacting). Moreover, besides reviewing the above-mentioned topics some further insight on the ghost issue and the infinite energy extrapolation is provided.  There is the hope that in the future this scenario might provide a phenomenologically viable and UV complete relativistic field theory of all interactions. 
  
  \vspace{0.9cm}

\noindent --------------------------------------------------------------------------------------------------------------------------------

\vspace{1.5cm}


\vspace{2cm}

\noindent Email: alberto.salvio@cern.ch


\newpage

\tableofcontents

\newpage

\section{Introduction and summary}

Relativistic field theories are the commonly accepted framework to describe  particle physics and gravity, at least at currently accessible energies. An important question is  whether such framework could hold up to infinite energies and still agree with the experimental data.
There are two serious difficulties that one has to overcome in order to give a positive answer to such a challenging question: the non-renormalizability of Einstein gravity~\cite{Goroff:1985sz,Goroff:1985th} and the presence of Landau poles in the Standard Model (SM). 

Even if one does not quantize the gravitational field, it is known that quantum corrections due to any relativistic  QFT generate terms that are not present in the Einstein-Hilbert action: specifically, local terms quadratic in the curvature tensor and with coefficients of dimension of non-negative powers of energy are generated~\cite{Utiyama:1962sn}, even if one sets them to zero at the classical level.  Therefore, it is not possible to avoid them in a relativistic field theory. The resulting theory is commonly known as quadratic gravity\footnote{Other names sometimes used  are ``$R^2$ gravity" and ``higher derivative gravity" as terms quadratic in the curvature have more than two derivatives of the gravitational field.} (QG). Starobinsky~\cite{Starobinsky:1980te} exploited these unavoidable terms and noted that a non-singular solution that is initially in the de Sitter space can be obtained by taking them into account. This resulted in a pioneering model of inflation, one of the models favoured by the Planck collaboration~\cite{Ade:2015lrj}.

What happens if the quantum dynamics of the gravitational field is taken into account in QG? Weinberg~\cite{Weinberg:1974tw} and Deser~\cite{Deser:1975nv} suggested that QG is renormalizable 
(all physical quantities can be made finite by redefining the parameters and re-normalizing  the fields) and few years later Stelle proved it rigorously~\cite{Stelle:1976gc}.

The presence of these local quadratic terms implies that classical QG belongs to the class of  higher derivative theories analysed long time ago by Ostrogradsky~\cite{ostro}, who proved that their Hamiltonian is unbounded from below. In QG this manifests in the presence of a massive ghost, which is the price to pay to have a relativistic field theory of quantum gravity\footnote{It should be noted that QG is distinct from the asymptotic safety proposal for quantum gravity made in~\cite{WeinbergAS}, where all possible terms (including the non-renormalizable ones beyond the quadratic order) are included: in QG only renormalizable interactions are introduced so that only a finite number of parameters are present. This {\it guarantees} the predictivity of the theory.
Possibly the ghost can be
avoided by introducing an infinite series of higher-derivative terms~\cite{Biswas:2005qr}, which can be viewed as non-local terms, but the resulting gravity theories 
contain infinite free parameters and are not known to be renormalizable.}. The importance of the quantum gravity problem has, however, encouraged several physicists to investigate whether QG can make sense and some recent progress in the ghost problem has been made. Most of the work done so far addressed the ghost problem within  finite dimensional quantum mechanical models and therefore the case of a relativistic field theory (and in particular of QG) remains an important target for future research.

Another potential issue of QG is  the clash between stability (understood as the absence of tachyons)  and the absence of Landau poles~\cite{Stelle:1977ry,Avramidi:1985ki}: whenever the parameters were chosen to ensure stability, perturbation theory featured a Landau pole; specifically, this Landau pole affected the parameter $f_0$ appearing in the Lagrangian as $\sqrt{-g} R^2/f_0^2$, where $g$ is the determinant of the spacetime metric $g_{\mu\nu}$ and $R$ is the Ricci scalar.  Some recent progress has also been made in this problem. In Ref.~\cite{Salvio:2017qkx} it was shown that QG coupled to a renormalizable QFT can hold up to infinite energies provided that all couplings flow to a UV fixed point and the gravitational sector flows to conformal gravity (a version of gravity that is invariant under
 Weyl transformations, $g_{\mu\nu}(x) \to  e^{2\sigma(x)} g_{\mu\nu}(x)$,
where $\sigma$ is a generic function of the spacetime point $x$.). The requirement that the QFT part enjoys a UV fixed point indicates the presence of several particles beyond the SM, which  could be searched for with current and/or future particle experiments and could account for the strong evidence of new physics that we undoubtedly already have (neutrino oscillations, dark mater, etc.).

The aim of this work is to review what is known so far about QG (taking into account the coupling to a general renormalizable QFT). Other monographs and books on QG are present in the literature (see e.g.~\cite{Avramidi:1986mj,Buchbinder:1992rb}, which focused on the renormalization of the theory). This review also includes  the recent progress on the two problems mentioned above (the ghost and the Landau poles) and provides further insight on these issues.
The article is structured as follows.
\begin{itemize}
\item  In Sec.~\ref{The theory} the action of QG coupled to a generic renormalizable QFT is discussed and the known physical degrees of freedom are identified with a new physically transparent method. 
\item Sec.~\ref{Renormalization} discusses the renormalizability of the theory; given that   detailed proofs are present   in the literature and, as mentioned above, books and reviews on this subject already exist, we recall and elucidate a known intuitive argument in favour of renormalizability by providing more details than those currently available.  In Sec.~\ref{Renormalization} we also collect from the existing literature the full  renormalization group equations (RGEs) for the dimensionless and dimensionful parameters of QG coupled to the most general renormalizable QFT. 
\item Sec.~\ref{Ghosts} is devoted to a pedagogical and detailed discussion of the ghost problem and the recent progress that has been made on this subject; most of the discussion, however, will be limited to simple finite dimensional quantum mechanical models and the extension to the full QG case remains an important target for future research. 
\item Sec.~\ref{Reaching infinite energy} reviews the issue of the Landau poles and how QG con flow to conformal gravity even in the presence of a generic QFT sector.
\end{itemize}

\section{The theory  (including a general matter sector)}\label{The theory}

In this review we do not consider only pure gravity, but also its couplings to a general renormalizable matter sector. 

\subsection{Jordan-frame Lagrangian}

The full action in the so-called Jordan frame is 
\be S=\int d^4x  \sqrt{-g} \, \Lag, \qquad \Lag=  \mathscr{L}_{\rm gravity} + \mathscr{L}_{\rm matter}+  \mathscr{L}_{\rm non-minimal} \label{TotAction}. \ee
We describe in turn the three pieces: the pure gravitational Lagrangian $\mathscr{L}_{\rm gravity}$, the matter Lagrangian $ \mathscr{L}_{\rm matter}$ and the non-minimal couplings $\mathscr{L}_{\rm non-minimal}.$
\subsubsection*{The pure gravitational Lagrangian}

$\mathscr{L}_{\rm gravity}$  in quadratic gravity is obtained from the Einstein-Hilbert action by adding all  possible local   terms quadratic in the curvature, whose coefficients have the dimensionality of non-negative powers of energy:
 \be  \mathscr{L}_{\rm gravity} = \alpha R^2  +\beta R_{\mu\nu}R^{\mu\nu}+\gamma R_{\mu\nu\rho\sigma}R^{\mu\nu\rho\sigma} 
 -\frac{\bp^2}{2} R - \Lambda, \label{Lgravity}
 \ee 
 where $R_{\mu\nu\rho\sigma}$, $R_{\mu\nu}$ and $R$ are the Riemann tensor, Ricci tensor and Ricci scalar, respectively\footnote{In this review we use the signature $\eta_{\mu\nu} = {\rm diag}(+1,-1,-1,-1)$ and define $$R_{\mu\nu\,\,\, \sigma}^{\quad \rho} \equiv \partial_{\mu} \Gamma_{\nu \, \sigma}^{\,\rho}- \partial_{\nu} \Gamma_{\mu \, \sigma}^{\,\rho} +  \Gamma_{\mu \, \tau}^{\,\rho}\Gamma_{\nu \,\sigma }^{\,\tau}- \Gamma_{\nu \, \tau}^{\,\rho}\Gamma_{\mu \,\sigma }^{\,\tau}, \quad \Gamma_{\mu \, \sigma}^{\,\rho} \equiv \frac12 g^{\rho \tau}\left(\partial_{\mu} g_{\sigma\tau}+\partial_{\sigma} g_{\mu\tau}-\partial_{\tau} g_{\mu\sigma}\right), \quad R_{\mu\nu} \equiv R_{\rho\mu\,\,\, \nu}^{\quad \rho},\quad R\equiv g^{\mu\nu}R_{\mu\nu}.$$}
 and the greek indices are raised and lowered with $g_{\mu\nu}$. Furthermore, $\alpha$, $\beta$ and $\gamma$ 
 are generic real coefficients. If the theory lives on a spacetime with boundaries one should also introduce in  $\mathscr{L}_{\rm gravity}$ a term proportional to $\Box R$, where $\Box$ is the covariant d'Alembertian,  in order to preserve renormalizability~\cite{deBerredoPeixoto:2004if,Shapiro:2008sf,Salles:2017xsr}; in the applications described in this review such term will not play any role and, therefore, will be neglected.
  Finally, $\bp$ and $\Lambda$ are the reduced Planck mass and the cosmological constant.
 
  One combination of the terms in (\ref{Lgravity})  is a total (covariant) derivative, the topological Gauss-Bonnet term:
 \beq   G\equiv  R_{\mu\nu\rho\sigma}R^{\mu\nu\rho\sigma} - 4 R_{\mu\nu} R^{\mu\nu} + R^2  =  \frac{1}{4}\epsilon^{\mu\nu\rho\sigma}
\epsilon_{\alpha\beta\gamma\delta}R_{\,\,\,\,\,\, \mu\nu}^{\alpha\beta} R_{\,\,\,\,\,\,\rho\sigma}^{\gamma\delta}= \mbox{divs.}, \label{Gdef}
 \eeq
where $\epsilon_{\mu\nu\rho\sigma}$ is the antisymmetric Levi-Civita tensor and ``divs" represents the covariant divergence of some current.  This total derivative does not contribute to the field equations and can be often  ignored. It is therefore convenient to write 
(\ref{Lgravity}) as 
 \be  \mathscr{L}_{\rm gravity} = (\alpha - \gamma) R^2  +(\beta +4\gamma) R_{\mu\nu}R^{\mu\nu}+\gamma G 
 -\frac{\bp^2}{2} R - \Lambda. \label{Lgravity2}
 \ee 

Furthermore, for reasons that will become apparent  when the degrees of freedom will be identified in Sec.~\ref{dof}, it is also convenient to express $R_{\mu\nu}R^{\mu\nu}$ in terms of $W^2\equiv W_{\mu\nu\rho\sigma}W^{\mu\nu\rho\sigma}$, where $W_{\mu\nu\rho\sigma}$ is the Weyl tensor
\beq W_{\mu\nu\alpha\beta} \equiv R_{\mu\nu\alpha\beta} + \frac{1}{2} (
g_{\mu\beta} R_{\nu\alpha} -g_{\mu\alpha} R_{\nu\beta}+ g_{\nu\alpha} R_{\mu\beta} - g_{\nu\beta}R_{\mu\alpha})
+\frac16 (g_{\mu\alpha}g_{\nu\beta}-g_{\nu\alpha}g_{\mu\beta} )R.
\eeq
One has
\beq  \frac12 W_{\mu\nu\rho\sigma}W^{\mu\nu\rho\sigma}  =  \frac12 R_{\mu\nu\rho\sigma}R^{\mu\nu\rho\sigma} - R_{\mu\nu}R^{\mu\nu} + \frac16 R^2, \eeq
which, together with the definition of $G$ in (\ref{Gdef}), gives
\be  R_{\mu\nu}R^{\mu\nu}   =  \frac{W^2}{2} + \frac{R^2}{3} - \frac{G}{2}.  \ee 
By inserting this expression of $R_{\mu\nu}R^{\mu\nu}$ in (\ref{Lgravity2}) one finds 
 \be    \boxed{\mathscr{L}_{\rm gravity} =  \frac{R^2}{6f_0^2}   - \frac{W^2}{2 f_2^2}-\epsilon G 
 -\frac{\bp^2}{2} R - \Lambda.} \label{Lgravity3}
 \ee 
 where 
 \be f_0^2 \equiv  \frac1{2\beta +2\gamma+6\alpha}, \quad f_2^2 \equiv -\frac{1}{\beta + 4\gamma}, \quad \epsilon \equiv \frac{\beta}{2}+\gamma.\ee
We have introduced the squares $f_0^2$ and $f_2^2$  because the absence of tachyonic instabilities  requires $f_0^2> 0 $ and $f_2^2 >0$ as we will see in Secs.~\ref{Einstein frame Lagrangian},~\ref{dof} and, in a more general context, in Sec.~\ref{Reaching infinite energy}.

 \subsubsection*{The matter Lagrangian}
 The general matter content of a renormalizable theory includes real scalars $\phi_a$, Weyl fermions $\psi_j$ and vectors $V^A_\mu$ (with field strength $F_{\mu\nu}^A$) and its Lagrangian is 
\bea \label{eq:Lmatter}
\Lag_{\rm matter} &=&  
- \frac14 (F_{\mu\nu}^A)^2 + \frac{D_\mu \phi_a \, D^\mu \phi_a}{2}  + \bar\psi_j i\slashed{D} \psi_j  - \frac12 (Y^a_{ij} \psi_i\psi_j \phi_a + \hbox{h.c.}) \nonumber\\
&& 
- \mathscr{V}(\phi) -\frac12 (M_{ij}\psi_i\psi_j+\hbox{h.c.}), \label{Lmatter}
\eea %
where 
\be \mathscr{V}(\phi) = \frac{m^2_{ab}}{2} \phi_a \phi_b + \frac{A_{abc}}{3!}\phi_a\phi_b\phi_c+ \frac{\lambda_{abcd}}{4!} \phi_a\phi_b\phi_c\phi_d, \ee
where all terms are contracted in a gauge-invariant way.
The covariant derivatives are\footnote{The spin-connection $ \omega^{ab}_\mu$ is defined as usual by 
$ \omega^{ab}_\mu = e^a_{\,\, \nu} \partial_\mu e^{b\nu} + e^a_{\,\, \rho}  \Gamma_{\mu \, \sigma}^{\,\rho}  e^{b\sigma}$
and $\gamma_{ab} \equiv \frac14 [\gamma_a, \gamma_b]$. } 
$$D_\mu \phi_a = \partial_\mu \phi_a+ i \theta^A_{ab} V^A_\mu \phi_b \qquad D_\mu\psi_j = \partial_\mu \psi_j + i t^A_{jk}V^A_\mu\psi_k + \frac12 \omega^{ab}_\mu \gamma_{ab}  \psi_j $$ 
The gauge couplings are contained in the matrices $\theta^A$ and $t^A$, which are the generators of the gauge group in the scalar and fermion representation respectively, while $Y^a_{ij}$ and $\lambda_{abcd}$ are the Yukawa  and  quartic couplings respectively. We have also added general renormalizable mass terms and cubic scalar interactions.
 Of course, for specific assignments of the gauge and global symmetries some of these parameters can vanish, but here we  keep a general expression.   

\subsubsection*{The non-minimal couplings}
$\mathscr{L}_{\rm non-minimal}$ represents the non-minimal couplings between the scalar fields $\phi_a$  and $R$:
\be \mathscr{L}_{\rm non-minimal} = -\frac12 \xi_{ab} \phi_a\phi_b R, \label{Lnonminimal}\ee
%
where all terms are contracted in a gauge-invariant way.
Non-minimal couplings are required by renormalizability: if they are omitted at the classical level, quantum corrections generate them (as we will see in Sec.~\ref{RGEs of the dimensionless parameters}).

\subsection{Einstein frame Lagrangian}\label{Einstein frame Lagrangian}

The action in the Jordan frame is most suited to address quantum aspects and to make contact with particle physics. However, when it comes to cosmological applications it is often better to express the gravitational part of the theory in a form closer to Einstein gravity~\cite{Kannike:2015apa,Salvio:2017xul}. This will also help us in identifying the degrees of freedom in Sec.~\ref{dof}. We now review how to obtain such a form of the theory and, in doing so, we shall neglect quantum corrections, which are best studied in the Jordan frame anyway.

 The non-standard $R^2$ term can be removed by adding to the Lagrangian the term
\be  - \sqrt{-g} \,   \frac{(R+3f_0^2 \chi/2)^2}{6f_0^2},\ee
where  $\chi$ is an auxiliary field. This Lagrangian vanishes once the $\chi$ EOM are used and
we are therefore free to add it to the total Lagrangian. However, this has the effect of modifying the non-minimal couplings: the term linear in $R$ in the Lagrangian now reads 
\be  -\frac12 \sqrt{-g} \,  f(\chi,\phi) R , \qquad  \quad f(\chi,\phi)\equiv\bp^2 + \xi_{ab} \phi^a\phi^b+\chi.\ee
In order to get rid of this remaining non-standard   term we perform a Weyl transformation: 
\be g_{\mu\nu} \rightarrow \frac{\bp^2}{f}g_{\mu\nu} , \qquad \phi^a \rightarrow \left(\frac{f}{\bp^2}\right)^{1/2} \phi^a, \qquad  \psi_j \rightarrow \left(\frac{f}{\bp^2}\right)^{3/4}  \psi_j, \qquad V_\mu^A \rightarrow  V_\mu^A. \label{ConfTransf}\ee
Now the Lagrangian can still be written as in (\ref{TotAction}), but with 
 \be    \mathscr{L}_{\rm gravity} =   - \frac{W^2}{2 f_2^2}-\frac{\bp^2}{2} R + \mbox{divs.}, \qquad \mathscr{L}_{\rm non-minimal}=0, \label{LgravityE}
 \ee 
\bea \label{eq:LmatterE}
\Lag_{\rm matter} &=&  
- \frac14 (F_{\mu\nu}^A)^2 + \bar\psi_j i\slashed{D} \psi_j  - \frac12 (Y^a_{ij} \psi_i\psi_j \phi_a + \hbox{h.c.})-\frac{\sqrt{6}\bp}{2 \zeta}  (M_{ij}\psi_i\psi_j+\hbox{h.c.}) \nonumber\\
&&+\frac{6\bp^2}{\zeta^2} \, \frac{D_\mu \phi_a \, D^\mu \phi_a  + \partial_\mu\,  \zeta \partial^\mu\zeta}{2} -  U(\zeta,\phi),
\eea
where we defined\footnote{Notice that in order for the metric redefinition in (\ref{ConfTransf}) to be regular one has to have $f>0$ and thus we can safely take the square root of $f$.} $\zeta\equiv\sqrt{6f}$ and
\be U(\zeta,\phi) \equiv \frac{36 \bp^4}{\zeta^4}\left[\mathscr{V}(\phi)+\Lambda +\frac{3f_0^2}{8} \left(\frac{\zeta^2}{6} - \bp^2 - \xi_{ab}\phi_a\phi_b\right)^2\right]. \label{Udef}\ee
In $ \mathscr{L}_{\rm gravity}$ we have not written explicitly the total derivatives as they typically do not play an important role in cosmology.
These total derivatives emerge when the Weyl transformation is applied to the two terms proportional to $\epsilon$  in (\ref{Lgravity3}).

The advantage of this form of the Lagrangian, known as the Einstein frame, is the absence of the non-minimal couplings and  the $R^2$ term. The latter has effectively been traded with the new scalar $\zeta$, which appears non-polynomially: the scalar kinetic terms are non-canonical and cannot be put in the canonical form with further field redefinitions given that the scalar field metric is not flat; moreover, the Einstein frame potential $U$ differs considerably from the Jordan-frame one, $V+\Lambda$. 
This result is a particular case of a more general theorem involving generic functions $f(R)$ of the Ricci scalar (for a review on $f(R)$ theories see e.g.~\cite{Sotiriou:2008rp} and references therein). Also, notice that the $W^2$ term is also present in the Einstein frame.

\begin{figure}[t]
\begin{center}
 $ \includegraphics[scale=0.6]{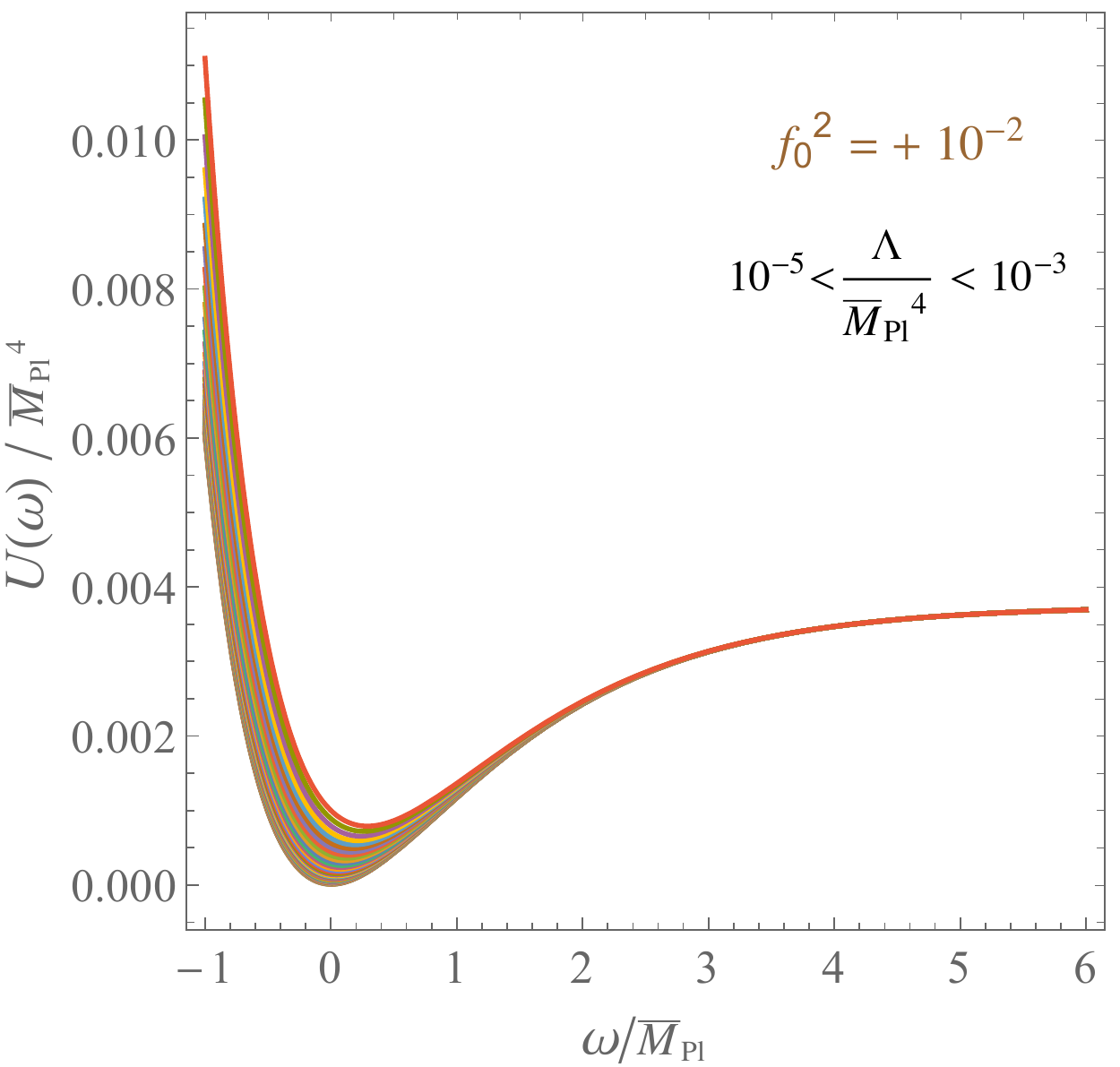}  \qquad \qquad  \includegraphics[scale=0.6]{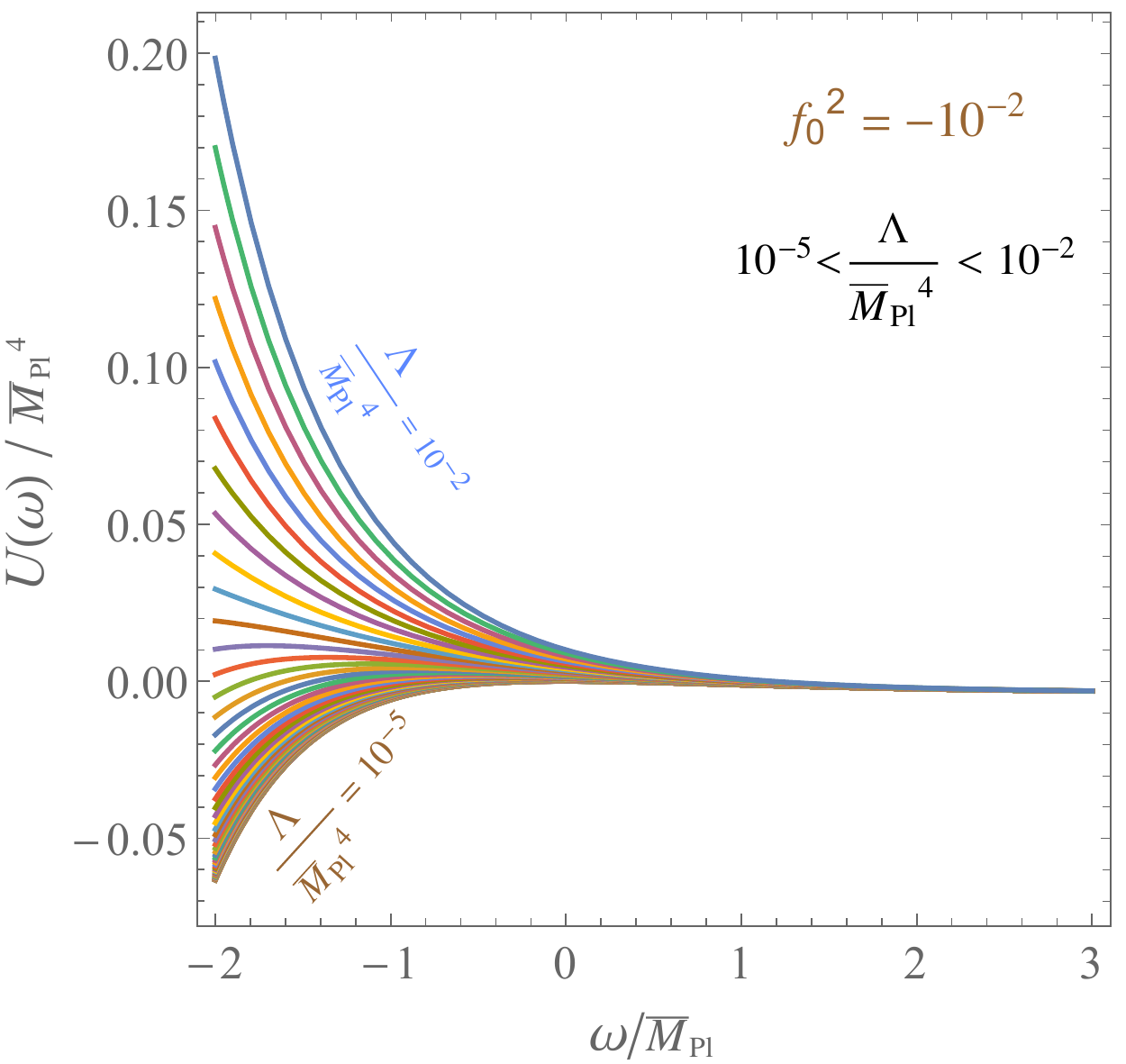} $
  \end{center}
   \caption{\em Einstein frame potential as a function of the canonically normalized scalar $\omega$ equivalent to the scalar $\zeta$ corresponding to the $R^2$ term in the Lagrangian. The quantity $f_0^2$ is chosen to be positive (negative) on the left (right). A minimum exists only for $f_0^2 >0$, which corresponds to Starobinsky's inflationary model.}
\label{Uomega}
\end{figure}

It is instructive to write the potential for $\zeta$ when the other fields $\phi_a$ are not present or are at the minimum of the potential and are not allowed to fluctuate (for example because they have very large masses). In this case one can make the kinetic term of $\zeta$ canonical through the field  redefinition $\zeta =\sqrt{6}\bp \exp(\omega/(\sqrt{6}\bp))$. The new field $\omega$ feels a potential 
\be U(\omega) =  \Lambda e^{-4 \omega/\sqrt{6}\bp} + \frac{3f_0^2 \bp^4}{8}\left(1-e^{-2\omega/\sqrt{6}\bp}\right)^2,\ee
where we have neglected $\mathscr{V}(\phi)$ and $\xi_{ab}\phi_a\phi_b$ as they can be absorbed in $\Lambda$ and $\bp^2$ when the scalar fields $\phi_a$ are absent or they are fixed to constant values. This is the potential of the famous Starobinsky's inflationary model~\cite{Starobinsky:1980te}. There is a stationary point of $U$ for
\be e^{-2\omega/\sqrt{6}\bp} = \frac{3f_0^2 \bp^4/8}{\Lambda + 3f_0^2 \bp^4/8} \ee
whenever the right-hand side of the equation above is positive. For positive cosmological constant, $\Lambda>0$, such stationary point always exists for $f_0^2 > 0$ when it is a point of minimum, but for $f_0^2<0$ either the stationary point does not exist or it is a point of maximum, not of minimum. This situation is illustrated in Fig.~\ref{Uomega} and it is a special case of a more general result (valid when the other scalars $\phi_a$ can fluctuate freely), which   proves that a minimum of the potential exists only for $f_0^2 > 0$ and will be presented in Sec.~\ref{Reaching infinite energy}.

\subsection{The degrees of freedom of quadratic gravity}\label{dof}
In Sec.~\ref{Einstein frame Lagrangian} we have seen that the $R^2$ term is  equivalent to a real scalar $\zeta$. We now complete the determination of the degrees of freedom of QG. We do so by working in the Einstein frame, where the gravity Lagrangian  is the one in (\ref{LgravityE}). The degrees of freedom associated with the matter Lagrangian can be identified with standard field theory methods and, therefore, we  do not discuss them explicitly here.

The total derivatives (``divs") in (\ref{LgravityE}) do not modify the degrees of freedom and for this reason will be neglected. Therefore, we focus on the following two terms in the gravity action:
\be S_W + S_{EH}, \ee
where $S_W$ is  the part due to the unusual $W$eyl-squared term,
\be S_W=\int d^4x  \sqrt{- g} \,\bigg[-  \frac{W^2}{2 f_2^2} \bigg], \label{action2}
 \ee
 and $S_{EH}$ is the usual $E$instein-$H$ilbert part,
\be S_{EH}=\int d^4x  \sqrt{-g} \,\bigg[ -\frac{\bp^2}{2}R
\bigg]. 
 \ee 

We will use a $3+1$ formalism (where space and time are treated separately). We do so because the identification of the degrees of freedom is  particularly simple within that formalism.

In this section, however, we will expand the metric around the flat spacetime, $ds^2_{\rm flat} = dt^2 - d\vec{x}^2$ as that is sufficient to determine the degrees of freedom 
perturbatively\footnote{For a discussion of a possible form of the non-perturbative spectrum see Refs.~\cite{Holdom:2015kbf,Holdom:2016xfn}.}.
By  choosing the Newtonian gauge, the metric describing the small linear fluctuations around the flat spacetime can be written as 
\be ds^2= (1+2\Phi(t, \vec{x})) dt^2 -2 V_i(t, \vec{x}) dt dx^i  - \left[(1-2\Psi(t, \vec{x})) \delta_{ij}+h_{ij}(t, \vec{x})\right]dx^idx^j. \label{dsPert}\ee 
By definition,  the vector  $V_i$ 
(not to be confused with the spatial components of the gauge fields $V_\mu^A$) and the tensor $h_{ij}$ perturbations satisfy
\be \partial_iV_i=0, \qquad h_{ij}=h_{ji},  \qquad h_{ii} =0, \qquad \partial_ih_{ij}=0. \label{Cond1}\ee
The Newtonian gauge is often used to study the small linear fluctuations around the Friedmann-Robertson-Walker (FRW) cosmological metric (see e.g.\cite{Weinberg:2008zzc} for a textbook treatment). Instead  we here study the fluctuations around the flat spacetime for simplicity. Also, sometimes the Newtonian gauge is defined for the scalar perturbations $\Phi$ and $\Psi$ only  (see e.g. \cite{Weinberg:2008zzc}). Here we consider a generalization, which also includes the non-scalar perturbations\footnote{A possible gauge dependent divergence of $h_{ij}$ has been set to zero by choosing the gauge appropriately.}. 
In Table \ref{table:dof} we provide the degrees of freedom of the gravitational sector (the part of the spectrum due to $\mathscr{L}_{\rm gravity}$). This includes the scalar $\zeta$ found in Sec.~\ref{Einstein frame Lagrangian} and the ordinary graviton and a massive spin-2 ghost graviton, which will be identified in the next sections (\ref{Helicity-2 sector},~\ref{Helicity-1 sector} and~\ref{Helicity-0 sector}).

\begin{table}[t]
\centering
\begin{tabular}{ |p{4cm}|p{4cm}|p{4cm}| }
\hline
\rowcolor{red!80!green!40!yellow!10}Field& spin & mass \\
\hline
 Graviton & 2 & 0 \\
Ghost &  2   & $M_2\equiv f_2 \bp/\sqrt{2}$ \\
Scalar $\zeta$ & 0 & $M_0\equiv f_0 \bp/\sqrt{2} + \dots$  \\
\hline
\end{tabular}
\caption{\it Degrees of freedom in the gravitational sector. The scalar $\zeta$ is due to the $R^2$ term in the Lagrangian; the dots in its mass $M_0$ represent the possible contribution of other scalars mixing with $\zeta$ (if any), which can be present in specific models.}
\label{table:dof}
\end{table}

\subsubsection{Helicity-2 sector}\label{Helicity-2 sector}
We start with the helicity-2 sector, whose quadratic action we denote with $S^{(2)}$. Both  $S_{EH}$ and $S_W$ contribute to this action: the helicity-2 quadratic action from $S_{EH}$ and $S_W$ are, respectively,
\bea S_{ES}^{(2)} &=& \frac{\bp^2}{8} \int d^4x \, \left(\dot h_{ij}\dot h_{ij}+h_{ij}\vec{\nabla}^2h_{ij}\right), \nonumber \\ 
S^{(2)}_W &=&- \frac{1}{4f_2^2} \int d^4x \left( \ddot h_{ij} \ddot h_{ij}+2 \dot h_{ij}\vec{\nabla}^2 \dot h_{ij} + h_{ij}\vec{\nabla}^4 h_{ij}\right),\eea
where a dot denotes a derivative w.r.t.~to time $t$,  $\vec{\nabla}^4 \equiv (\vec{\nabla}^2)^2$ and $\vec{\nabla}^2$ is the three-dimensional Laplacian.
 Thus
\be S^{(2)} = S_{EH}+ S_W =  \frac{\bp^2}{8M_2^2} \int d^4x \left[-     \ddot h_{ij} \ddot h_{ij}-2 \dot h_{ij}\vec{\nabla}^2 \dot h_{ij} - h_{ij}\vec{\nabla}^4 h_{ij}+M_2^2 \left(\dot h_{ij}\dot h_{ij}+h_{ij}\vec{\nabla}^2h_{ij}\right)\right], \ee 
where $M_2^2 \equiv f_2^2 \bp^2/2$.

One can go to momentum space with a  spatial Fourier transform
\bea h_{ij}(t, \vec{x}) =  \int \frac{d^3p}{(2\pi)^{3/2}}  e^{i\vec{p}\cdot \vec{x}} \sum_{\lambda= \pm 2}  h_\lambda(t,\vec{p}) e^\lambda_{ij} (\hat p),  \label{ExpMom}  \eea
where $e^\lambda_{ij} (\hat p)$ are the usual polarization tensors for helicities $\lambda= \pm 2$. We recall that for $\hat p$ along the third axis the polarization tensors satisfying (\ref{Cond1}) are  given by 
\be e^{+2}_{11} = -e^{+2}_{22} = 1/2, \quad e^{+2}_{12} = e^{+2}_{21} = i/2, \quad e^{+2}_{3i} =e^{+2}_{i3} = 0, \quad  e^{-2}_{ij} =  (e^{+2}_{ij})^* \label{PolT}\ee 
and for a generic momentum direction $\hat p$ we can obtain $e^\lambda_{ij} (\hat p)$ by applying to (\ref{PolT}) the rotation that connects the third axis with  $\hat p$. The polarization tensors defined in this way also obey the orthonormality condition
\be e^\lambda_{ij} (\hat q) (e^{\lambda'}_{ij} (\hat q))^* = \delta^{\lambda\lambda'}.  \label{PolOrt}\ee 
By using the Fourier expansion in (\ref{ExpMom}) one obtains
\bea S^{(2)} &=&  \frac{\bp^2}{8M_2^2}  \sum_{\lambda= \pm 2} \int dt d^3p  \left[-     \ddot h_\lambda^* \ddot h_\lambda+2 \dot h_\lambda^*\vec{p}^{\, \, 2} \dot h_\lambda - h_\lambda^*\vec{p}^{\,\,4} h_\lambda+M_2^2 \left(\dot h_\lambda^*\dot h_\lambda-h_\lambda^*\vec{p}^{\, \, 2}h_\lambda\right)\right] \nonumber \\
&=& \frac{\bp^2}{8M_2^2}  \sum_{\lambda= \pm 2} \int dt d^3p  \left[-     \ddot h_\lambda^* \ddot h_\lambda+(\omega_1^2+\omega_2^2) |\dot h_\lambda|^2 - \omega_1^2\omega_2^2  |h_\lambda|^2\right], \label{S2helicity2} \eea 
where 
\be  \omega_1\equiv  \sqrt{\vec{p}^{\, \, 2} + M_2^2}, \qquad  \omega_2\equiv |\vec{p}\,|. \label{omega12QG} \ee 
The action $S^{(2)}$ is the sum of the actions of Pais-Uhlenbeck oscillators, which will be studied in Sec.~\ref{The Pais-Uhlenbeck model}. There we will see  that this system is equivalent to a ghost  d.o.f. with frequency $\omega_1$ and a normal  d.o.f. with frequency $\omega_2$. Therefore, the conclusion is  that the helicity-2 sector features a massless field (the ordinary graviton) and a ghost field\footnote{Starting from the initial action~(\ref{Lgravity3}), it is possible to perform field redefinitions and use the auxiliary field method  to make the ghost field explicitly appear in the Lagrangian~\cite{Hindawi:1995an}. This is equivalent to what has been done in Sec.~\ref{Einstein frame Lagrangian} to make the scalar field $\zeta$ appear explicitly in the Lagrangian.} with mass $M_2\equiv f_2 \bp/\sqrt{2}$.  Thus, as anticipated before, we see that $f_2^2>0$ is required to avoid tachyonic instabilities. Lorentz invariance implies that the helicity-1 and helicity-0 components of the massive ghost should be present too. We will see how they emerge in the next Secs.~\ref{Helicity-1 sector} and~\ref{Helicity-0 sector}. The derivation of the ghost field presented here simplifies and agrees with previous proofs based on the $h_{\mu\nu}$ propagator~\cite{Stelle:1976gc,Johnston:1987ue}.

\subsubsection{Helicity-1 sector}\label{Helicity-1 sector}

 Next, we move to the helicity-1 sector, whose quadratic action is denoted here with $S^{(1)}$. $S^{(1)}$ is given by the sum of the Einstein-Hilbert contribution 
 \be S_{EH}^{(1)} = \frac{\bp^2}{4} \int d^4x \, \left(\partial_i V_{j}\right)^2,\ee
and the Weyl contribution
 \be S^{(1)}_W =- \frac{1}{2f_2^2} \int d^4x \left(\partial_i \dot V_j \partial_i \dot V_j - V_i \vec{\nabla}^4 V_i\right). \ee
 Thus the full quadratic action in the helicity-1 sector is 
\be S^{(1)} = \int d^4x  \frac{\bp^2}{4M_2^2}\left[ \dot V_j \vec{\nabla}^2 \dot V_j + V_i \vec{\nabla}^4 V_i-M_2^2V_{j}\vec{\nabla}^2 V_{j} \right].  \label{LagV}\ee 
Given that $\vec{\nabla}^2$ is a negatively-defined operator, we see that $V_i$ has a ghost kinetic term and a mass $M_2$ and has therefore to be identified with the helicity-1 components of the massive spin-2 ghost.

\subsubsection{Helicity-0 sector}\label{Helicity-0 sector}
We denote the helicity-0 action with $S^{(0)}$, which has one contribution from the Weyl-squared term and one from the Einstein-Hilbert term,  $S^{(0)}=S^{(0)}_W+S^{(0)}_{EH}$.  Expanding around the flat spacetime  leads to the following helicity-0 action (modulo total derivatives)
\bea S^{(0)}_W &=& -\frac{2}{3f_2^2}\int d^4 x \left[\vec{\nabla}^2\left(\Phi+\Psi\right) \right]^2, \label{SW} \\ 
S^{(S)}_{ES} &=&\frac{\bp^2}{2}\int d^4 x \left[-6 \dot\Psi^2 +4 \Psi \vec{\nabla}^2 \Phi -2 \Psi \vec{\nabla}^2 \Psi  \right]. \label{SES}\eea

 The variation of $S^{(0)}$ with respect to $\Phi$ gives
  \be -\frac{4}{3 f_2^2 \bp^2} \vec{\nabla}^4 \left(\Phi+\Psi\right) +2\vec{\nabla}^2 \Psi  = 0.\label{ConstraintPhi}\ee
  We see that this equation does not depend on the time derivative of the fields and, therefore, has to be considered as a constraint. Solving for $\Phi$:
 \be \Phi = - \Psi +3 M_2^2\vec{\nabla}^{-2} \Psi. \label{PhiConstr} \ee 
 In the expression above $\vec{\nabla}^{-2}$ denotes the inverse Laplacian, which can be defined by going to momentum space, $\vec{p}$, and identifying $\vec{\nabla}^{-2} \to - 1/\vec{p}^{\, \, 2}$. Inserting (\ref{PhiConstr}) into Eqs.~(\ref{SW}) and (\ref{SES}) gives 
 \be S^{(0)} = \frac{\bp^2}{2}\int d^4 x \left[-6 \dot\Psi^2 -6 \Psi \vec{\nabla}^2 \Psi + 6M_2^2 \Psi^2  \right]  = 3\bp^2\int d^4 x \left[ -(\partial\Psi)^2 + M_2^2 \Psi^2 \right]. \label{S0action}\ee 
 We see that the kinetic term of $\Psi$ is of the ghost type and its mass is $M_2$. Therefore, $\Psi$ represents the helicity-0 component of the ghost spin-2 field.

\section{Renormalization}\label{Renormalization}
One of the main motivations for considering QG is its improved quantum behaviour with respect to Einstein theory. Therefore, it seems appropriate to discuss the renormalization properties right after the definition of the theory.  

\subsection{Renormalizability}\label{Renormalizability}

The renormalizability of QG is suggested by simple power counting arguments, general covariance  and dimensional analysis, it is therefore not surprising that some authors~\cite{Weinberg:1974tw,Deser:1975nv} noted this property several decades ago. There are also formal proofs~\cite{Stelle:1976gc,Barvinsky:2017zlx} of the renormalizability of QG, but we do not reproduce them here because they are described in detail in the original articles\footnote{These formal derivations can also be extended to include the general renormalizable matter sector considered in Sec.~\ref{The theory}.}.

 It is illuminating, however, to recall the main ingredients of the intuitive arguments in favour of renormalizability. Let us consider the expansion of QG around the flat spacetime, $g_{\mu\nu}= \eta_{\mu\nu}+h_{\mu\nu}$, and a generic loop correction in momentum space.  The vertices involving $h_{\mu\nu}$ contain at most 4 powers of the momenta $p$, while the $h_{\mu\nu}$-propagator  behaves as $1/p^4$ for large  momenta if an appropriate quantization is used~\cite{Stelle:1976gc} (see below). Therefore, in this case, the superficial degree of divergence should  be four or less (see, for example, Chapter 12 of~\cite{Weinberg:1995mt}). This conclusion holds both in the pure QG and in the presence of the most general renormalizable QFT.

It is instructive to illustrate the quantization  that leads to a propagator that behaves as $1/p^4$ for large  momenta. The presence of the ordinary graviton and the spin-2 ghost with mass $M_2$ tells us that the $h_{\mu\nu}$-propagator should have two poles,
\be \frac{Z_{\rm graviton}}{p^2 + i \epsilon}, \qquad \frac{Z_{\rm ghost}}{p^2-M_2^2 +i\epsilon'},\ee 
where $Z_{\rm graviton}$ and $Z_{\rm ghost}$ are the corresponding residues and we have allowed for two a priori different prescriptions, $\epsilon$ and $\epsilon'$.   The two poles 
are both proportional to the same tensor structure as they both have spin-2.
  The requirement that the $h_{\mu\nu}$-propagator behaves as $p^4$ for large momenta leads to the condition $Z_{\rm graviton} = - Z_{\rm ghost}$.
  In this case  
  the  $h_{\mu\nu}$-propagator is proportional to
\be
\frac{1}{p^2 + i \epsilon}-\frac{1}{p^2 -M_2^2+ i \epsilon'} =
- \frac{M_2^2}{(p^2+i\epsilon)(p^2-M_2^2+i\epsilon)} +\pi i  \delta(p^2-M_2^2) ({\rm sign}(\epsilon')-{\rm sign}(\epsilon)), \label{grPmghP}
\ee
where we have used the formula
\be \frac1{x\pm i \epsilon} = \mathscr{P}\frac1{x} \mp i \pi \delta(x)\ee
with $\mathscr{P}$ being the principal part.
The second term on the right-hand side of Eq.~(\ref{grPmghP}) corresponds to the fact that the poles are shifted in different directions in the complex energy plane for ${\rm sign}(\epsilon')\neq{\rm sign}(\epsilon)$. Therefore,
one obtains a propagator that behaves as $1/p^4$ only if\footnote{To convince ourselves of the correctness of this statement one could  insert the propagator in~(\ref{grPmghP}) in a loop integral; the effect of the Dirac $\delta$-function is to drop one momentum integration and to add a power of momentum at the denominator, for a total of two (not four) momenta in the power counting.}
${\rm sign}(\epsilon')={\rm sign}(\epsilon)$. Given that the absolute values of $\epsilon$ and $\epsilon'$ are not important this final condition can be simplified to $\epsilon = \epsilon'$.

The condition $\epsilon = \epsilon'$ implies that the ghost should be quantized by introducing an indefinite metric on the Hilbert space~\cite{Stelle:1976gc}. The easiest way to show this is by looking at the action $S^{(0)}$ of the helicity-0 component of the ghost in (\ref{S0action}), this allows us to avoid the complications due  to spacetime indices. The corresponding Lagrangian is
\be \Lag^{(0)} = \frac{1}{2}\left( - \dot\Psi^2 - \Psi \vec{\nabla}^2 \Psi + M_2^2 \Psi^2\right), \ee
where we have canonically normalized $\Psi$ by rescaling $\Psi \to \Psi/\sqrt{6}\bp$. The conjugate variable is then
\be \Pi_\Psi = \frac{\partial\Lag^{(0)}}{\partial\dot\Psi}  = - \dot\Psi \ee
and the canonical commutators are
\be [\Psi(t,\vec{x}), \dot\Psi(t,\vec{y})] = - i \delta^{(3)}(\vec{x} - \vec{y}), \quad [\Psi(t,\vec{x}), \Psi(t,\vec{y})] = 0, \quad [\dot\Psi(t,\vec{x}), \dot\Psi(t,\vec{y})] = 0.  \label{canCommPsi} \ee
Performing a spatial Fourier transform and demanding $\Psi$ to solve its EOM leads to
\be  \Psi(t,\vec{x}) =  \int \frac{d^3p}{\sqrt{2(2\pi)^{3}\omega(\vec{p})}} \left(b_0(\vec{p}) e^{i \vec{p}\cdot\vec{x}-i \omega(\vec{p}) t} +b_0(\vec{p})^\dagger e^{-i \vec{p}\cdot\vec{x}+i \omega(\vec{p}) t} \right), \ee
where $\omega(\vec{p})\equiv \sqrt{\vec{p}^{\,\, 2}+ M_2^2}$,
and the commutation rules above imply
  \be [ b_0(\vec{p}),  b_0(\vec{q})^\dagger] = -  \delta(\vec{p}-\vec{q}), \qquad [ b_0(\vec{p}),  b_0(\vec{q})] = 0.  \label{bComm}\ee
At this point we have a choice: we can 
\begin{description}
\item[1] interpret the $b_0 \, (b_0^\dagger)$ as annihilation (creation) operators, 
\item[2] interpret the $b_0 \, (b_0^\dagger)$ as creation (annihilation) operators.
\end{description} 
 In  Case 1, as we will see in Sec.~\ref{The Pais-Uhlenbeck model: quantum}, one should introduce an indefinite metric on the Hilbert space; in Case 2 the  indefinite metric can be avoided, but the energies are negative:
 this statement will be shown in Sec.~\ref{The Pais-Uhlenbeck model: quantum}, but its correctness is  intuitive because
  in that case one would interpret $-\omega(\vec{p})$ (rather than $+\omega(\vec{p})$) as the energy. Let us compute the propagator $P(x)$  in the two cases. The definition is 
 \be P(x)\equiv \langle 0|T\Psi(t,\vec{x})\Psi(0) | 0 \rangle = \theta(t) P_+(x) + \theta(-t)P_-(x),   \ee
 where
\be P_+(x) \equiv \langle 0|\Psi(t,\vec{x})\Psi(0) | 0 \rangle, \,\, P_+(x) \equiv \langle 0|\Psi(0) \Psi(t,\vec{x}) | 0 \rangle \ee 

\begin{description}
\item[1] In Case 1 we have 
\be P_+(x) =  - \int \frac{d^3p \,\, e^{-i px}}{2(2\pi)^3 p_0}, \qquad P_-(x) = P_+(-x)  \label{P+1}\ee
where $p_0\equiv \omega(\vec{p})$. The minus sign in (\ref{P+1}) is  due to the minus sign in the commutation relation~(\ref{bComm}). Therefore, by using a standard text-book derivation,
\be P(x) = - \int \frac{d^4p \, e^{-ipx}}{(2\pi)^4 (p^2 - M_2^2 + i \epsilon)},\ee
where $\epsilon >0$. We see that this corresponds to $Z_{\rm ghost} =  - Z_{\rm graviton}$ and $\epsilon' = \epsilon$.
\item[2] In Case 2 we still have
\be P_+(x) =  - \int \frac{d^3p \,\, e^{-i px}}{2(2\pi)^3 p_0}, \qquad P_-(x) = P_+(-x),  \label{P+2}\ee
but now $p_0 = - \omega(\vec{p})$ (the energies are negative) and one ends up with 
\be P(x) = - \int \frac{d^4p \, e^{-ipx}}{(2\pi)^4 (p^2 - M_2^2 - i \epsilon)}.\ee
Note that the overall minus sign has a different origin than in Case 1: here it is due to the negative energy condition $p_0 = - \omega(\vec{p})$, not to the commutators as the role of $b_0$ and $b_0^\dagger$ is switched. So in this case one still has $Z_{\rm ghost} =  - Z_{\rm graviton}$ but $\epsilon' =- \epsilon$ and renormalizability does not occur.
\end{description}
Therefore, the conclusion is  that renormalizability requires a quantization with an indefinite metric on the Hilbert space. In Sec.~\ref{The Pais-Uhlenbeck model: quantum} we will show that such a metric should be introduced also to ensure that the Hamiltonian  is bounded from below. This raises an interpretational problem as in quantum mechanics the positivity of the metric is related to the positivity of probabilities. This problem will be addressed in Sec.~\ref{Probabilities}, where the state of the art of the related literature will be discussed.

\subsection{RGEs}

The renormalizability of the theory (including the gravitational sector) allows us to use the standard renormalization group machinery developed for field theories without gravity. The modified minimal subtraction  ($\overline{\rm MS}$) scheme will be adopted in this review. 

\subsubsection{RGEs of the dimensionless parameters}\label{RGEs of the dimensionless parameters}
The 1-loop RGEs of the dimensionless parameters are independent of the dimensionful quantities and it is thus convenient to present them separately. Their expression for a general renormalizable matter sector is 
{\allowdisplaybreaks\bea\label{RGGadim}
 \frac{df_2^2}{d\tau}&=& -f_2^4\bigg(\frac{133}{10} +\frac{N_V}{5}+\frac{N_F}{20}+\frac{N_S}{60}
\bigg),\\
 \frac{df_0^2}{d\tau}&=&  \frac53 f_2^4 + 5 f_2^2 f_0^2 + \frac56 f_0^4 +\frac{f_0^4}{12} (\delta_{ab}+6\xi_{ab})(\delta_{ab}+6\xi_{ab}), \label{RGGadim2}\\
 \frac{d\epsilon}{d\tau} &=& -\left[\frac{196}{45} +\frac1{360}\left(62N_V + \frac{11}{2} N_F + N_S\right) \right],  \label{RGGadim3}\\
 \frac{d\xi_{ab}}{d\tau}  &=& \frac16\lambda_{abcd}\left(6\xi_{cd}+\delta_{cd}\right) 
+ (6\xi_{ab}+\delta_{ab})   \sum_{k=a,b} \left[\frac{Y_2^k}{6} - \frac{C_{2S}^k}{2}  \right]+\nonumber \\
&& -\frac{5f_2^4}{3f_0^2} \xi_{ab}
+f_0^2  \xi_{ac}\left( \xi_{cd}+\frac23 \delta_{cd}\right)(6 \xi_{db}+\delta_{db}),\label{eq:RGExi} \\
 \frac{dY^a}{d\tau} &=& \frac12(Y^{\dagger b}Y^b Y^a + Y^a Y^{\dagger b}Y^b)+ 2 Y^b Y^{\dagger a} Y^b + \nonumber\\
&&+ Y^b {\rm \, Tr}(Y^{\dagger b} Y^a) 
- 3 \{ C_{2F} , Y^a\}  + \frac{15}{8}f_2^2 Y^a, \\
 \frac{d\lambda_{abcd}}{d\tau} &=&  \sum_{\rm perms} \bigg[\frac18
  \lambda_{abef}\lambda_{efcd}+
  \frac38 \{\theta^A,\theta^B\}_{ab}\{\theta^A ,\theta^B\}_{cd}
  -{\rm Tr}\, Y^a Y^{\dagger b} Y^c Y^{\dagger d}+
   \nonumber
\\
&&\label{eq:RGElambda}
+\frac58 f_2^4 \xi_{ab}\xi_{cd}+  \frac{ f_0^4}{8}\xi_{ae}\xi_{cf}(\delta_{eb}+6\xi_{eb})(\delta_{fd}+6\xi_{fd}) +\\
&&+\frac{f_0^2}{4!}  (\delta_{ae}+6\xi_{ae})(\delta_{bf}+6\xi_{bf})\lambda_{efcd}\bigg]
+ \lambda_{abcd} \bigg[ \sum_{k=a,b,c,d} (Y_2^k-3  C_{2S}^k)+ 5 f_2^2\bigg]   ,\nonumber
\eea}
where 
\be \tau\equiv \ln\left(\mu/\mu_0\right)/(4\pi)^2,\ee
 $\mu$ is the $\overline{\rm MS}$ energy scale and $\mu_0$ is a fixed energy,  $N_V$, $N_F$ and $N_S$ are the numbers of gauge fields, Weyl fermions and real scalars. Also, $Y_2^k$, $C_{2S}^k$ and $C_{2F}$ are defined  by 
\be {\rm Tr} (Y^{\dagger a}Y^b) =Y_2^a \delta ^{ab}, \quad \theta^A_{ac} \theta^A_{cb}= C_{2S}^{a} \delta_{ab}, \quad C_{2F} = t^A t^A.\ee 
The sum over ``perms" in the RGEs of the $\lambda_{abcd}$ runs over the $4!$ permutations of $abcd$.  We do not show  the RGEs of the gauge couplings because they are not modified by the gravitational couplings (see \cite{Fradkin:1981iu} and~\cite{Narain:2012te,Narain:2013eea,Salvio:2014soa}).

Some terms in the 2-loop RGEs have been determined~\cite{Salvio:2017qkx}. For example, switching off all couplings but $f_0$ one obtains the 2-loop RGE for $f_0$~\cite{Salvio:2017qkx}
\be \frac{d f_0^2}{d\tau} =  
\frac{5}{6} f_0^4  -  \frac{1}{(4\pi)^2} \frac{5}{12} f_0^6  .
\ee
However, a complete expression of the 2-loop RGEs for all couplings is not available yet.

Note that the coefficient $\epsilon$ of the topological term $G$ does not appear in the RGEs of the other parameters. Indeed, $G$  
vanishes when the spacetime is topologically equivalent to the flat spacetime and the RGEs, being UV effects, are independent of the global spacetime properties.

The RGEs above are the result of several works. The first attempt to determine the RGEs of $f_2$ and $f_0$ was presented in Ref.~\cite{Julve:1978xn}. The results of \cite{Julve:1978xn} are incomplete and contain some errors. An improved calculation was later provided by~\cite{Fradkin:1981hx,Fradkin:1981iu}, which, however, still contains an error in the RGE of $f_0$. The first correct calculation of the RGE of $f_0$ in the pure gravity case appeared in~\cite{Avramidi:1985ki}; indeed, the result of~\cite{Avramidi:1985ki} was later checked by~\cite{Codello:2006in,Salvio:2014soa} with completely different techniques. Ref.~\cite{Salvio:2014soa}  also extended the results of~\cite{Avramidi:1985ki} to include the general couplings to renormalizable matter sectors. 
The RGE for $\epsilon$ in the presence of general renormalizable matter fields can be found in~\cite{Avramidi:1986mj} (see also~\cite{Einhorn:2014bka} for a more recent discussion). Also, Ref.~\cite{Ohta:2013uca} checked the RGEs of $f_2$, $f_0$ and $\epsilon$ with functional renormalization group methods.

Eqs.~(\ref{RGGadim})-(\ref{RGGadim2}) clearly show that even if the spacetime metric is not quantized and we do not introduce the terms quadratic in the curvature in the Lagrangian, such terms are anyhow generated by loops of matter fields, as originally showed in~\cite{Utiyama:1962sn}.

\subsubsection{RGEs of the dimensionful  parameters}

The 1-loop RGEs of the dimensionful  parameters are
{\allowdisplaybreaks\bea \label{sys:RGm} 
\frac{d  \bp^2}{d\tau}  &=& \frac{1}{3} m_{aa}^2+\frac{1}{3}{\rm \,Tr}(M^\dagger M) +2\xi_{ab} m_{ab}^2  +\left(\frac{2f_0^2}{3}-\frac{5f_2^4}{3f_0^2}+2X\right)\bp^2,\label{RGEPlanck}\\
\frac{d  \Lambda}{d\tau}  &=&\frac{m_{ab}^2 m_{ab}^2}{2}-
{\rm \,Tr}[(MM^\dagger)^2]+\frac{5f_2^4+f_0^4}{8} \bp^4+(5f_2^2+f_0^2) \Lambda+4\Lambda X,\\
 \frac{dM}{d\tau}  &=& \frac12(Y^{\dagger b}Y^b M + M Y^{\dagger b}Y^b)+ 2Y^b M^{\dagger } Y^b +  Y^b {\rm \,Tr}(Y^{\dagger b} M)+ \nonumber\\
&& - 3 \{ C_{2F} , M\}  + \frac{15}{8}f_2^2 M+MX,\\
\frac{d m^2_{ab}  }{d\tau} &=&\lambda_{abef} m_{ef}^2+ A_{aef}A_{bef} -2
  \big[{\rm \,Tr}(Y^{\{a} Y^{\dagger b\}} M M^{\dagger})+\nonumber \\ && 
 +{\rm \,Tr}( Y^{\dagger \{a} Y^{b\}}  M^\dagger M)
+{\rm \,Tr}\, (Y^a M^{\dagger }Y^b M^{\dagger})+  {\rm \,Tr}\, (MY^{\dagger a} MY^{\dagger b} )\big]+\nonumber\\ && \nonumber+\frac52 f_2^4 \xi_{ab} \bp^2+\frac{f_0^4}{2}\left(\xi_{ab}+6\xi_{ae}\xi_{eb}\right) \bp^2+ \\
 &&+f_0^2 \left(m_{ab}^2 +3\xi_{bf} m_{af}^2+3\xi_{af}m_{bf}^2 +6 \xi_{ae}\xi_{bf} m_{ef}^2\right)+\nonumber\\ &&+m_{ab}^2 \left[\sum_{k=a,b} (Y_2^k-3  C_{2S}^k)+ 5 f_2^2+2X\right], \\
\frac{d A_{abc}  }{d\tau} &=& \lambda_{abef}A_{efc}+ \lambda_{acef}A_{efb}+ \lambda_{bcef}A_{efa}+ \nonumber \\
&&   -2{\rm \,Tr}\left(Y^{\{a}Y^{\dagger b} Y^{c\}}M^\dagger\right)- 2{\rm \,Tr}\left(Y^{\dagger\{c}Y^{a} Y^{\dagger b\}}M\right)+\nonumber \\&&+f_0^2\left(A_{abc}+3\xi_{af}A_{fbc}+3\xi_{bf}A_{fac}+3\xi_{cf}A_{fab}\right)+ \nonumber\\&&+6f_0^2\left(\xi_{ae}\xi_{bf}A_{efc}+\xi_{ae}\xi_{cf}A_{efb}+\xi_{be}\xi_{cf}A_{efa}\right)+\nonumber \\  && + A_{abc}\left[\sum_{k=a,b,c} (Y_2^k-3  C_{2S}^k)+ 5 f_2^2+X\right],  \label{RGEcubic}
\eea}
where the curly brackets represent the sum over the permutations of the corresponding indices: e.g. $Y^{\{a} Y^{\dagger b\}}= Y^{a} Y^{\dagger b}+Y^{b} Y^{\dagger a}$.
The symbol $X$ represents a gauge-dependent quantity~\cite{Salvio:2017qkx}.
 The RGEs of massive parameters are gauge dependent as the unit of mass is gauge dependent.
Any dimensionless ratio of dimensionful parameters is physical and the corresponding RGE is indeed gauge-independent,
as it can be easily checked from Eqs.~(\ref{RGEPlanck})-(\ref{RGEcubic}).

The RGEs above for the most general renormalizable matter sector where obtained in Ref.~\cite{Salvio:2017qkx} and later checked in Ref.~\cite{Anselmi:2018ibi}. However, before~\cite{Salvio:2017qkx} appeared, a number of articles computed the RGEs of some massive parameters in less general models. The RGE for $\Lambda/\bp^4$ in the pure gravity theory was determined in~\cite{Avramidi:1985ki} and a detailed description of the methods used can be found in~\cite{Avramidi:1986mj}. The RGE of the ratio between the Higgs squared mass $M_h^2$ and $\bp^2$ was computed in~\cite{Salvio:2014soa} (where the matter sector was identified with the SM).

These general RGEs can be used to address issues related to the high-energy extrapolation, such as the the UV-completeness or the vacuum stability of generic theories of the sort studied here. 

\section{Ghosts}\label{Ghosts}

In this section we discuss systems (such as quadratic gravity) featuring ghosts, recall the related problems and present some possible solutions. We will mostly focus on finite dimensional systems, but discuss both classical and quantum mechanical aspects.

 \subsection{Ghosts in classical mechanics}\label{Lagrange and Hamilton}

We consider  a physical system described by a certain number of coordinates\footnote{Note that the case of fields can be obtained by interpreting the index $i$ as a space coordinate $\vec{x}$.} $q_i$ and restrict our attention to  Lagrangians that depend on $q_i$, $\dot q_i$, $\ddot q_i$ and, possibly, on time $t$,
\be L(q, \dot q, \ddot q, t), \ee 
where the dot is the derivative w.r.t.~$t$
and, from now on, we understand the index $i$.
 This setup covers the case we are interested in: the Lagrangian of quadratic gravity depends both on the first and second derivatives of the field variables 
 because of the extra terms quadratic in the curvature; moreover, an explicit dependence on time emerges e.g. when a cosmological background is considered~\cite{Salvio:2017xul}.  
 
In the following paragraphs we will first discuss the derivation of Euler-Lagrange equations of motion and then introduce the Hamiltonian approach. 
 This discussion will be valid for QG as a particular case.

The least action principle in this context tells us that the variation $\delta S$ of the action  $S\equiv \int dt L$ with respect to variations $\delta q$ of the coordinates that vanish on the time boundaries (together with their first derivatives, $\delta\dot q$) should be zero\footnote{The summation on the index $i$ is understood: for example $\frac{\partial L}{\partial q} \delta q \equiv \sum_i \frac{\partial L}{\partial q_i} \delta q_i$.}: 
\be 0 = \delta S = \int dt \left(\frac{\partial L}{\partial q} \delta q + \frac{\partial L}{\partial \dot q} \delta \dot q +\frac{\partial L}{\partial \ddot q} \delta \ddot q\right). \label{deltaS}\ee
Here we should require that also $\delta\dot q$ vanishes on the time boundaries because the values of $q$ at two times are not sufficient to identify the motion as the equations   involve derivatives higher than the second  order.
By integrating by parts the second term in (\ref{deltaS}) once and the third term twice we obtain the Euler-Lagrange equations of motion for four-derivative theories:
\be \frac{d}{dt} \left(\frac{\partial L}{\partial \dot q} - \frac{d}{dt} \frac{\partial L}{\partial \ddot q} \right) = \frac{\partial L}{\partial q}. \ee 

We now move to the Hamiltonian approach. We start by defining two canonical coordinates 
\be q_1 \equiv q, \qquad q_2 \equiv \dot q. \label{qlDef} \ee
In this case  the conjugate momenta are defined by
\be p_l \equiv \frac{\delta  L}{\delta \dot q_l} \equiv \frac{\partial  L}{\partial \dot q_l} -\frac{d}{dt}\frac{\partial L}{\partial \ddot q_l}, \label{plDef} \ee
where the index $l$ runs over $\{1,2\}$. A motivation for this definition will be given below in Sec.~\ref{The Ostrogradsky theorem}. For $l=1$ and $l=2$ separately the conjugate momenta read
\be p_1 =   \frac{\partial  L}{\partial \dot q} -\frac{d}{dt}\frac{\partial L}{\partial \ddot q}, \qquad p_2 =   \frac{\partial  L}{\partial \ddot q}. \label{simplepDef}\ee
Then, one defines as usual the Hamiltonian $H$ as 
\be H = p_l \dot q_l- L(q, \dot q, \ddot q, t).  \label{Hdef}\ee

\subsubsection{The Ostrogradsky theorem}\label{The Ostrogradsky theorem}

 Under a non-degeneracy assumption, i.e. the fact that\footnote{$\partial^2 L/\partial \ddot q^2$ denotes   the Hessian matrix of $L$, whose elements are $\partial^2 L/\partial \ddot q_i\partial \ddot q_j$.} $\det(\partial^2 L/\partial \ddot q^2)\neq 0$, it is possible to argue that the system is classically unstable\footnote{Lagrangians that depend on even higher derivatives of $q$ have been considered in the literature in the time-independent case~\cite{Pais}, but these situations go beyond our scope as the quadratic gravity Lagrangian only depends on the derivative of $q$ up to the second order.
}. 
 
Indeed, this assumption allows us to express  $\ddot q$ as 
\be \ddot q = f(q, \dot q, p_2, t), \label{qdd}\ee 
where $f$ is the inverse of $\partial L/\partial\ddot q$ viewed as a function of $\ddot q$. 
  Once Eqs.~(\ref{qlDef}) and~(\ref{qdd}) are used, $H$ reads
\be H = p_1 q_2 +p_2 f(q_1,q_2,p_2, t) - L(q_1,q_2,f(q_1,q_2,p_2, t),t), \label{Hostro} \ee
which is manifestly a function of the form
\be H = H(q_l, p_l, t).  \label{Hdep}\ee

The form of $H$ in (\ref{Hostro}) implies the celebrated {\it Ostrogradsky theorem}~\cite{ostro}: {\it the Hamiltonian obtained from a Lagrangian of the form $L(q, \dot q, \ddot q, t)$, which depends non-degenerately on $\ddot q$ (i.e. $\det(\partial^2 L/\partial \ddot q^2)\neq 0$), is not bounded from below.} Indeed, the expression of $H$ in (\ref{Hostro}) shows that $H$ depends linearly on the momentum $p_1$ and therefore goes to $-\infty$ if $p_1$ tends either to $+\infty$ or $-\infty$ (when $q_2$ is non-vanishing).
Note that this result is  valid for QG as a particular case.

One may wonder why the conjugate momenta is defined as in (\ref{plDef}). The reason is that the standard form of the Hamiltonian equations of motion follows in this case and, therefore, the Hamiltonian is a constant of motion if it does not depend explicitly on time. In order to see this let us consider an infinitesimal variation of the Hamiltonian and  compute it in two different ways, by using (\ref{Hdef}) and (\ref{Hdep}). Respectively we have 
\bea dH &=&   p_l d\dot q_l + \dot q_l dp_l -  \frac{\partial L}{\partial q} d q - \frac{\partial L}{\partial \dot q} d \dot q -\frac{\partial L}{\partial \ddot q}d \ddot q - \frac{\partial L}{\partial t}dt ,  \\ 
dH &=& \frac{\partial H}{\partial q_l} d q_l + \frac{\partial H}{\partial p_l} d p_l +  \frac{\partial H}{\partial t}dt.  \label{dH2}\eea
By using  the definition of the conjugate momenta in (\ref{simplepDef}) and $q_2 = \dot q$ 
in the first expression of $dH$ we obtain
\be dH = \dot q_l dp_l  -  \frac{\partial L}{\partial q} d q -\frac{d}{dt}  \frac{\partial L}{\partial \ddot q} d\dot q - \frac{\partial L}{\partial t}dt=\dot q_l dp_l  -  \frac{\partial L}{\partial q} d q - \dot p_2 d\dot q - \frac{\partial L}{\partial t}dt. \ee
The Euler-Lagrange equations allow us to write the term $ \frac{\partial L}{\partial q} d q$ as follows 
\be  \frac{\partial L}{\partial q} d q  =   \frac{d}{dt} \left(\frac{\partial L}{\partial \dot q} - \frac{d}{dt} \frac{\partial L}{\partial \ddot q} \right) dq = \dot p_1 dq \ee
so 
\be dH = \dot q_l dp_l  - \dot p_l dq_l  - \frac{\partial L}{\partial t}dt.  \ee 
By comparing now this expression with the one in (\ref{dH2}) we obtain 
\be \dot q_l = \frac{\partial H}{\partial p_l}, \qquad \dot p_l = -  \frac{\partial H}{\partial q_l}, \qquad   \frac{\partial H}{\partial t}= -\frac{\partial L}{\partial t}. \label{HamEOMs} \ee
Therefore we see that in theories with a Lagrangian of the form $L(q, \dot q, \ddot q, t)$, which depends non-degenerately on $\ddot q$ (i.e. $\det(\partial^2 L/\partial \ddot q^2)\neq 0$), the Hamiltonian equations have the standard form provided that the definition of the conjugate momenta are modified according to (\ref{plDef}). By inserting the first two equations in (\ref{HamEOMs}) into (\ref{dH2}) we obtain that the Hamiltonian is a constant of motion provided that $\partial H/\partial t=  0$.

\subsubsection*{(In)stabilities}

If a system fulfills the hypothesis of the Ostrogradsky theorem it can develop instabilities. However, this theorem does not directly imply that all solutions of such a system are unstable. Here by ``stable solution" we mean a solution of the equations of motion such that for initial conditions close enough to the region of the phase space spanned by this solution the motion is bounded (it does not run away). There are several examples of systems of this type that feature bounded motions: the Pais-Uhlenbeck model~\cite{Pais} to be discussed in Sec.~\ref{The Pais-Uhlenbeck model}  (in some cases even in the presence of interactions~\cite{Pagani:1987ue,Smilga:2004cy,Pavsic:2013noa,Kaparulin:2014vpa,Pavsic:2016ykq,Smilga:2017arl}) and quadratic gravity expanded at linear level around the flat or de Sitter spacetime~\cite{Ivanov:2016hcm,Tokareva:2016ied,Salvio:2017xul}.

\subsubsection{The Pais-Uhlenbeck model}\label{The Pais-Uhlenbeck model}

The Ostrogradsky theorem applies to a large class of higher derivative theories, but we have seen that it does not forbid directly the existence of stable solutions. To understand further the issues of higher derivative theories it is convenient to analyse a simple system, which captures some of the essential characteristics of quadratic gravity. In this section we therefore focus on the Pais-Uhlenbeck model~\cite{Pais}, whose 
Lagrangian is 
\beq \label{eq:LagO}
L =  -\frac{\ddot q^2}{2} + (\omega_1^2+\omega_2^2) \frac{\dot q^2}{2} - \omega_1^2 \omega_2^2 \frac{q^2}{2}-V(q)
=-\frac12  q (\frac{d^2}{dt^2} + \omega_1^2)(\frac{d^2}{dt^2} + \omega_2^2)q - V(q) + \mbox{ total derivatives}.
\eeq
Here $V$ is a function of $q$ representing a possible interaction and $\omega_1$ and $\omega_2$ are real parameters. As we will see, $\omega_1$ and $\omega_2$ represent the frequencies of two decoupled oscillators when $V=0$.  Apart from its simplicity, another reason for considering this model is that it closely resembles the helicity-2 sector of QG (see Eq.~(\ref{S2helicity2})). In QG $\omega_1\neq \omega_2$ at finite spatial momentum (see (\ref{omega12QG})); therefore, the unequal frequency case is particularly relevant. 

\subsubsection*{\it Lagrangian analysis}
 
The Lagrangian equation of motion is 
\beq \label{eq:classq''''}
(\frac{d^2}{dt^2} + \omega_1^2)(\frac{d^2}{dt^2} + \omega_2^2)q+V'(q)=
\frac{d^4 q}{dt^4}+ (\omega_1^2 +\omega^2_2)\frac{d^2q}{dt^2} + \omega_1^2 \omega_2^2 q+V'(q)=0.\eeq
Eq.~(\ref{eq:classq''''}) makes it manifest why one chooses $\omega_1^2$ and $\omega_2^2$ to be positive: otherwise the solutions of the equations of motion would feature tachyonic instabilities at least for vanishing $V$.  

The corresponding classical solution,
for given initial conditions $q_0\equiv q(0), \dot q_0\equiv\dot q(0), \ddot q_0\equiv\ddot q(0), \dddot{q}_0\equiv \dddot q(0)$ at $t=0$,
is
\beq \label{eq:q(t)}
q(t) = -\frac{\omega_2^2 q_0 + \ddot q_0}{\omega_1^2-\omega_2^2} \cos(\omega_1 t) + 
\frac{\omega_1^2 q_0 + \ddot q_0}{\omega_1^2-\omega_2^2} \cos(\omega_2 t) -
\frac{\omega_2^2 \dot q_0 + \dddot q_0}{\omega_1(\omega_1^2-\omega_2^2)} \sin(\omega_1 t) +
\frac{\omega_1^2 \dot q_0 + \dddot q_0}{\omega_2(\omega_1^2-\omega_2^2)} \sin(\omega_2 t) .
\eeq
This is a well behaved system without run-away issues for unequal frequencies, $\omega_1 \neq \omega_2$.
By taking the limit $\omega_1 \rightarrow \omega_2\equiv \omega$ in the expression above one obtains 
\beq \label{eq:q2(t)}
q(t) = \sin  (t \omega ) \left[\frac{t \left(q_0 \omega ^2+\ddot q_0\right)}{2 \omega }+\frac{3 \dot q_0 \omega ^2+\dddot q_0}{2 \omega ^3}\right]+\cos  (t \omega ) \left[q_0-\frac{t \left(\dot q_0 \omega ^2+\dddot q_0\right)}{2 \omega ^2}\right].
\eeq
Note that the amplitudes of the sine and cosine functions above grow linearly with $t$.

\begin{figure}[t]
\begin{center}
  \includegraphics[scale=0.6]{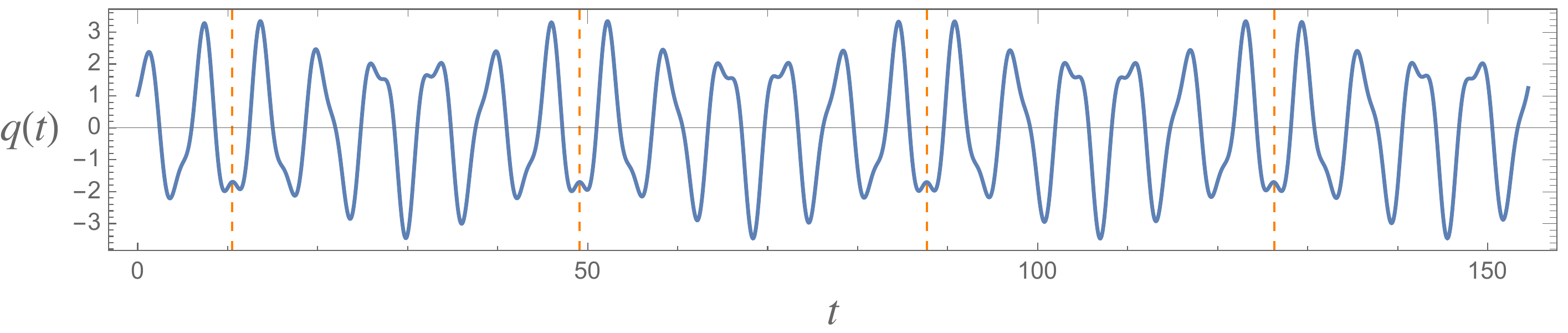} 
  \end{center}
   \caption{\em Solution to the equation of motion (\ref{eq:classq''''}) of the Pais-Ulhenbeck model with $V(q)= \lambda \sin(q)^4$. The plot is presented in units of $\omega_2$. The other parameters are set as follows: $\omega_1 = 2.1 $, $\lambda =1.022$. The motion appears to be bounded and periodic (the vertical dashed lines indicate the period).}
\label{sinq}
\end{figure}

Run-away (i.e.~unstable) solutions can also appear for $\omega_1 \neq \omega_2$ if a non-quadratic potential, i.e.~$V\neq 0$, is introduced. However, it has been found numerically that  the system admits stable solutions regardless of the unboundedness of the Hamiltonian for some choices of $V$, such as $V(q)\propto \sin(q)^4$ \cite{Pavsic:2013noa}. The situation for this potential is illustrated in Fig.~\ref{sinq}. In~\cite{Pavsic:2016ykq} it was found that the solutions are unstable unless $V$ is bounded from below {\it and} above. Of course, this can only be generically true for $\omega_1 \neq \omega_2$ because, for equal frequencies, we have seen that the motion is unbounded even for $V =0$, which is certainly bounded from  below and above.

\subsubsection*{\it Hamiltonian analysis}

We can now construct the Hamiltonian\footnote{An analogous construction for QG was performed in~\cite{Buchbinder:1987vp,Buchbinder:1987av,Kluson:2013hza,Buchbinder:1992rb}.} by using the general formul$\ae$ of Sec.~\ref{Lagrange and Hamilton}. Ostrogradsky's canonical variables defined in (\ref{plDef}) and~(\ref{qlDef}) in this case read

\beq \label{eq:Ostroq2}
\begin{array}{ll}
q_1 = q,\qquad  &\displaystyle 
p_1 =  \frac{\partial  L}{\partial \dot q} -\frac{d}{dt}\frac{\partial L}{\partial \ddot q}  =(\omega_1^2+\omega_2^2)\dot q + \dddot q ,\\[5mm]
q_2 = \dot q,&\displaystyle p_2 = \frac{\partial  L}{\partial \ddot q} = -\ddot q.
\end{array}\eeq
Note that the non-degeneracy hypothesis of the Ostrogradsky theorem is obviously satisfied in this case: $\partial^2 L/\partial \ddot q^2 = -1 \neq 0$.
 Indeed, by using the general formula in (\ref{Hostro}) we obtain (in the Pais-Uhlenbeck model $f(q,\dot q, p_2,t)=-p_2$)
\beq \label{eq:HOstro}
H = p_1 q_2 - \frac{1}{2}p_2^2 - \frac{\omega_1^2+\omega_2^2}{2} q_2^2 +\frac{\omega_1^2\omega_2^2}{2}q_1^2+V(q_1), 
 \eeq
which is obviously unbounded from below. 
From (\ref{HamEOMs}) the Hamiltonian equations of motion are
\beq \left\{\begin{array}{ll}\displaystyle
\dot q_1 =   \frac{\partial H}{\partial p_1} =q_2,\qquad  &\displaystyle 
\dot p_1 =  -\frac{\partial H}{\partial q_1}=  -\omega_1^2\omega_2^2 q_1 - V'(q_1) ,\\[5mm]  \displaystyle
\dot q_2 =   \frac{\partial H}{\partial p_2}=-  p_2,&\displaystyle 
\dot p_2 =   -\frac{\partial H}{\partial q_2}=- p_1  + (\omega_1^2+\omega_2^2) q_2.
\end{array}\right.\eeq
They imply the classical Euler-Lagrange equation of motion in (\ref{eq:classq''''}).
 
 When $\omega_1\neq \omega_2$, the Hamiltonian in~(\ref{eq:HOstro}) can be brought in diagonal form
 (except for the effect of the interaction $V$)
\beq H = -\frac12 (\tilde p^{ 2}_1  +\omega_1^2 \tilde q_1^{ 2}) + \frac12 (\tilde p^{2}_2 + \omega_2^2 \tilde q^{ 2}_2) + V(q_1) \label{DecH1}\eeq
through the canonical transformation
\beq \label{eq:canqp}
q_1 =\frac{ \tilde{q}_2 -   \tilde{p}_1/\omega_1}{\sqrt{\omega_1^2 - \omega_2^2}},\qquad
{q_2} = \frac{\tilde{p}_2 - \omega_1\tilde{q}_1}{\sqrt{\omega_1^2 - \omega_2^2}},\qquad
p_1 = \omega_1\frac{\omega_1 \tilde{p}_2 -\omega_2^2\tilde{q}_1}{\sqrt{\omega_1^2 - \omega_2^2}},\qquad
p_2  = \frac{\omega_2^2 \tilde{q}_2-\omega_1 \tilde{p}_1 }{\sqrt{\omega_1^2 - \omega_2^2}}.\eeq
which satisfies $q_1 p_1 - q_2 p_2 = \tilde{p}_2 \tilde{q}_2-\tilde{p}_1\tilde{q}_1$. 
Its inverse is
\beq  \label{eq:canqp'}
\tilde{q}_1 =\frac{p_1 - \omega_1^2q_2}{\omega_1\sqrt{\omega_1^2-\omega_2^2}},
\qquad
\tilde{q}_2 = \frac{\omega_1^2q_1  - p_2 }{\sqrt{\omega_1^2-\omega_2^2}},\qquad
 \tilde{p}_1 = \omega_1\frac{\omega_2^2 q_1  - p_2}{\sqrt{\omega_1^2-\omega_2^2}},\qquad
\tilde{p}_2 =\frac{p_1 - \omega_2^2q_2}{\sqrt{\omega_1^2-\omega_2^2}}
.\eeq
Note that, given the first equation in~(\ref{eq:canqp}), $V(q_1)$ introduces interactions between $\tilde q_2$ and $\tilde p_1$. However, from (\ref{DecH1}) one can see that the system for $V=0$ is equivalent to two decoupled oscillators with frequencies $\omega_1$ and $\omega_2$. Note that the first oscillator contributes negatively to the Hamiltonian: this is the manifestation of the Ostrogradsky theorem in this basis. Since the derivation of  (\ref{DecH1}) is valid only for $\omega_1\neq \omega_2$ (because otherwise the transformation in (\ref{eq:canqp}) would be singular) one might hope to have a classical Hamiltonian that is bounded from below for $\omega_1=\omega_2$. This is not the case as the Hamiltonian in the form given in (\ref{eq:HOstro}) is valid for $\omega_1=\omega_2$ too and is not bounded from below.

\subsection{Quantum mechanics with ghosts}

Before examining the peculiar features of the quantization with ghosts, let us spell out some basic assumptions of standard quantum mechanics, which will be made in the presence of ghosts too,  
including in the case of QG.

\begin{itemize}
\item Quantizing the theory consists in substituting  the canonical coordinates $q_j$ and conjugate momenta $p_j$ with some operators acting on a vector space, whose elements are identified with the possible states of the system\footnote{For simplicity, in the following we will use the same symbol to denote the quantum operators and the corresponding classical variables (when this does not create confusion).}.
\item The Hamiltonian $H$ in quantum mechanics is defined as a self-adjoint operator ($H^\dagger = H$) with respect to some metric on the vector space of states. $H$ generates the time evolution: the state $|\psi_t\rangle$ at time $t$ is given by 
\be |\psi_t\rangle = U(t) |\psi_0\rangle,  \qquad U(t) \equiv e^{-i H t}. \ee
Moreover, the Hamiltonian is assumed to have the same expression in terms of $q_j$ and $p_j$ as in classical mechanics, Eq.~(\ref{Hdep}).
\item The canonical coordinates $q_j$ and their conjugate momenta $p_j$ are promoted to operators by imposing the canonical commutators, i.e. 
\be [q_j,p_k] = i \delta_{jk}, \qquad  [q_j,q_k] = 0, \qquad  [p_j,p_k] = 0 \label{can-comm} \ee 
and requiring them to be self-adjoint: $q_j^\dagger=q_j$ and $p_j^\dagger = p_j$. 
\end{itemize}
Possible probabilistic interpretations of quantum theories with ghosts will be discussed in Sec.~\ref{Probabilities}.
 
Most of the efforts that have been done so far in quantizing theories with ghosts have focused on simple toy models, which isolate the main source of concern: the presence of four time-derivatives. The model that is typically studied is the quantum version of the Pais-Uhlenbeck construction given in Sec.~\ref{The Pais-Uhlenbeck model}, which is perhaps the simplest four-derivative extension of an ordinary quantum mechanical model. Therefore, we will mostly focus on it. However, some of the results reviewed in this section can be applied to other models too.

\subsubsection{Trading negative energies with negative norms}\label{The Pais-Uhlenbeck model: quantum}

A first thing one can prove is that some Hamiltonians that are not bounded from below can be quantized in a way that their quantum spectrum is instead bounded from below, but this is achieved by introducing an indefinite metric on the Hilbert space (as we will see, this is precisely the metric with respect to which $H$, $q_j$ and $p_j$ have been assumed to be self-adjoint). A classic example is the Pais-Uhlenbeck Hamiltonian\footnote{It is important to recall that Hamiltonian (\ref{DecH1}) is equivalent to the original Hamiltonian in (\ref{eq:HOstro}) when $\omega_1\neq \omega_2$ a condition that is assumed to hold here (for the quantization of the equal frequency Pais-Uhlenbeck model see e.g.~\cite{Mannheim:2004qz,Smilga:2005gb,Mannheim:2006rd,Bolonek2006,Smilga:2017arl})} in Eq.~(\ref{DecH1}) for vanishing $V$, which we now discuss in some detail. 

The part of the classical Hamiltonian that contributes negatively is 
\be H_1 \equiv -\frac12 (\tilde p^{ 2}_1  +\omega_1^2 \tilde q_1^{ 2}),\label{H1} \ee
and it is on this part that we shall focus as the other one $H_2\equiv \frac12 (\tilde p^{ 2}_2  +\omega_2^2 \tilde q_2^{ 2})$, being positive, can be quantized with standard methods. Note that the quadratic Hamiltonian of the ghost of QG can be written as the sum of Hamiltonians of the form (\ref{H1}), as is clear from Eqs.~(\ref{S0action}) and (\ref{LagV}) and the fact that the Lagrangian (\ref{S2helicity2}) of the helicity-2 sector of QG is the sum of Pais-Uhlenbeck Lagrangians.

What allows us to trade the negative energy in Eq.~(\ref{H1}) with negative norm is the exchange of creation and annihilation operators: one defines the annihilation and creation operators respectively as 
\be \tilde a_1 \equiv \sqrt{\frac{\omega_1}{2}}\left(\tilde q_1-i\frac{\tilde p_1}{\omega_1}\right), \qquad  
\tilde a_1^\dagger \equiv \sqrt{\frac{\omega_1}{2}}\left(\tilde q_1+i\frac{\tilde p_1}{\omega_1}\right), \label{annihilation-creation} \ee
where we used $\tilde q_1^\dagger = \tilde q_1$ and $\tilde p_1^\dagger = \tilde p_1$.
The relative signs between $\tilde q_1$ and $\tilde p_1$ have been switched with respect to~the standard case. We keep here the label $1$ to recall that the oscillator with label 2 is subject to the usual definition of annihilation and creation operators:
\be \tilde a_2 \equiv \sqrt{\frac{\omega_2}{2}}\left(\tilde q_2+i\frac{\tilde p_2}{\omega_2}\right), \qquad  
\tilde a_2^\dagger \equiv \sqrt{\frac{\omega_2}{2}}\left(\tilde q_2-i\frac{\tilde p_2}{\omega_2}\right). \label{annihilation-creation} \ee
 From the canonical commutators (\ref{can-comm}) and by using the canonical transformation in (\ref{eq:canqp'})  it follows
 \be [\tilde q_j,\tilde p_k] = i \delta_{jk}, \qquad  [\tilde q_j,\tilde q_k] = 0, \qquad  [\tilde p_j,\tilde p_k] = 0, \label{can-comm-tilde} \ee 
 which leads to
\be [\tilde a_j,\tilde  a_k^\dagger]=\eta_{jk}, \qquad [\tilde a_j, \tilde a_k]=0, \qquad [\tilde a^\dagger_j, \tilde a_k^\dagger]=0,\label{gen-comm}\ee
where $\eta_{11}=-1$, $\eta_{22}=1$, $\eta_{12}=\eta_{21}=0$.
One can now express $\tilde q_1$ and $\tilde p_1$ in terms of $\tilde a_1$ and $\tilde a_1^\dagger$  as usual and find
\be H_1=  -\omega_1 \tilde a_1^\dagger \tilde a_1 +\frac{\omega_1}{2}\equiv \omega_1 N_1 +\frac{\omega_1}{2} , \label{H1p}\ee
where we defined a number operator  $N_1\equiv -\tilde a_1^\dagger \tilde a_1$ (see below) with an unusual minus sign. Indeed, with this definition $N_1$, $\tilde a_1$ and $\tilde a_1^\dagger$ satisfy the usual commutation relations
\be [N_1, \tilde a_1] =-  \tilde a_1, \qquad [N_1, \tilde a_1^\dagger] =  \tilde a_1^\dagger, \ee
which allows 
us to interpret $\tilde a_1$ and $\tilde a_1^\dagger$ as annihilation and creation operators respectively: the eigenstates of $N_1$, i.e. $N_1|n_1\rangle=n_1|n_1\rangle$, satisfy
\be \tilde a_1 |n_1\rangle = c(n_1) |n_1- 1\rangle,\qquad \tilde a_1^\dagger |n_1\rangle = d(n_1) |n_1+ 1\rangle. \ee
$c$ and $d$ can be determined up to an overall phase, once the normalization of $ |n_1\rangle$ is fixed. Here, for reasons that will become clear shortly, we allow some norms to be negative and we choose the normalizations\footnote{More general assignments, $\nu_{n_1}\neq \pm 1$ are equivalent because we can always re-normalize the states in a way that $\nu_{n_1} =  \pm 1$ as long as there are no zero norm states, which we assume here.} $\langle n_1 |n_1\rangle =\nu_{n_1}$, where $\nu_{n_1}=\pm 1$. Notice now
\be -\nu_{n_1}n_1 = \langle n_1 | \tilde a_1^\dagger \tilde a_1 | n_1\rangle =|c(n_1)|^2 \langle n_1 -1 |n_1-1\rangle =|c(n_1)|^2\nu_{n_1-1},\ee
which leads to 
\be |c(n_1)|^2 =-\frac{\nu_{n_1}}{\nu_{n_1-1}}n_1. \label{cn} \ee
If all norms are positive, i.e. all $\nu_{n_1}=1$, it is possible to show with a standard textbook argument that the spectrum of $N_1$ (and therefore, because of Eq.~(\ref{H1p}), that of the Hamiltonian) is not bounded from below. This is because Eq.~(\ref{cn}) tells us $n_1<0$ and we can then reach an arbitrary large and negative value of $n_1$ by acting with the annihilation operator.

The only way to avoid $n_1<0$ is to take
$\nu_{n_1}=-\nu_{n_1-1}$. Indeed, in this case (\ref{cn}) gives\footnote{In order to fix $d(n_1)$ consider
 \be - \nu_{n_1}(n_1+1) =  \langle n_1 | a_1^\dagger  a_1 -1 | n_1\rangle = \langle n_1 | a_1 a_1^\dagger  | n_1\rangle =|d(n_{1})|^2 \langle n_1 +1 |n_1+1\rangle =|d(n_1)|^2\eta_{n_1 +1},\ee
 which gives
 \be |d(n_1)|^2 = n_1 + 1.\ee} 
\be |c(n_1)|^2 = n_1, \ee
which as usual implies that the spectrum of $N_1$ is $\{n_1\}=\{0,1,2,3,...\}$ (and therefore $N_1$ can be appropriately be identified with a number operator) and  the spectrum of the Hamiltonian is thus bounded from below. The state with $n_1 =0$ is interpreted as that without ghost quanta and so we require it to have positive norm. Therefore, $\nu_{n_1}=-\nu_{n_1-1}$ implies that the states with an even (odd) number of ghost quanta have positive (negative) norm.

A similar reasoning can be done in QG linearized  around the flat spacetime: the energy becomes bounded from below if an indefinite metric on the Hilbert space is introduced (See Sec.~\ref{Renormalizability}). Furthermore, we saw in Sec.~\ref{Renormalizability} that an indefinite metric should be present also in order for QG to be renormalizable. Therefore, insisting in having arbitrarily negative energies to preserve the positivity of the metric appears to have very little motivation.

As mentioned before, in this construction $q_j$, $p_j$ and $H$ are self-adjoint w.r.t.~the indefinite metric. This leads to problems in the definition of probabilities, which we shall address in  Sec.~\ref{Probabilities}.

\subsubsection{The problem of the wave-function normalization}
 
 So far we have given some features of the quantum theory, but we have not yet specified completely the quantization procedure. We still have to define the spectrum of the operators $q_j$. 
 
 Let us discuss this point in the Pais-Uhlenbeck model with $\omega_1\neq \omega_2$ for the sake of definiteness. One possibility would be to assume, as usual, that the spectrum is real for both $q_1$ and $q_2$. However, this leads to non-normalizable wave functions~\cite{Woodard:2006nt,Woodard:2015zca}. To see this, we consider the ground-state wave function  $\psi_{ 0}(q_1,q_2) \equiv \langle q_1,q_2|0\rangle$, where $|0\rangle$ is the vacuum, defined as  $\tilde a_1|0\rangle=0$ and $\tilde a_2|0\rangle=0$, while $|q_1,q_2\rangle$ is an eigenstate of $q_1$ and $q_2$.
Using the standard representation for the conjugate momentum acting on the wave functions, $p_i = - i \partial/\partial q_i$, one obtains the ground-state wave function 
\beq\label{eq:psi0div}
 \psi_{ 0}(q_1,q_2) \propto \exp\bigg(\frac{-q_1^2 \omega_1 \omega_2 +   q_2^2}{2}  (\omega_1 + \omega_2)  - i  q_1 q_2\omega_1 \omega_2\bigg).\eeq
 With this quantization, $\psi_{ 0}(q_1,q_2)$ is non-normalizable along the $q_2$-direction. However, $\psi_{ 0}(q_1,q_2)$ becomes normalizable when one performs the integral of $|\psi_{ 0}(q_1,q_2)|^2$ on the imaginary $q_2$-axis. 
 
 This suggests that one could obtain a consistent quantization by requiring $q_2$ to have a purely imaginary spectrum, while assuming a standard quantization (with real spectrum) for $q_1$~\cite{Salvio:2015gsi}.

 \subsubsection{The Dirac-Pauli quantization}
 
 The quantization with purely imaginary eigenvalues for a canonical variable $\hat x$ was first discussed by Pauli~\cite{Pauli} for Lagrangians with at most 2 time-derivatives, elaborating on a previous work by Dirac~\cite{Dirac}. In the rest of this work we will refer to this unusual quantization as the Dirac-Pauli  quantization.
 To proceed, let us deduce some basic properties of the Dirac-Pauli quantization for a generic variable $\hat x$. 
 
 The defining property is that the spectrum of $\hat x$ is purely imaginary:  
 \be \hat x|x\rangle = ix|x\rangle.\ee  It follows
 $\langle x'|\hat x|x\rangle = i x \langle x'|x\rangle$, which, together with the self-adjointness of $\hat x$, i.e. $\langle x'|\hat x|x\rangle = \langle x|\hat x|x'\rangle^*  = - i x' \langle x|x'\rangle^*= - i x' \langle x'|x\rangle$, implies
 $ (x+x')\langle x'|x\rangle  = 0.$
 The general solution to this equation is $\langle x'|x\rangle = \delta(x+x') h(x)$, where $h$ is a function that we set to 1 without loss of generality: this can always be done by rescaling the states $|x\rangle$. Then, one obtains
 \be  \langle x'|x\rangle = \delta(x+x') \label{xx'inner}\ee 
and the completeness\footnote{We require the completeness of the states $|x\rangle$ as part of the definition of the vector space.} condition reads
 \be \int dx |x\rangle \langle -x| = 1, \quad \iff \quad \int dx |x\rangle \langle x| = \eta, \quad \iff \quad \int dx |x\rangle \langle x|\eta =1, 
 \label{identityDP} \ee
 where $\eta$ is the operator defined by $\eta |x\rangle = |-x\rangle$.
 
 It can be shown that the variable $\hat p$ canonically conjugate to $\hat x$ is also a Dirac-Pauli variable: i.e. $\hat p|p\rangle = i p |p \rangle$ where $p$ is a generic real number. To show this we first notice that the operator $\exp(\hat p a)$, where $a$ is a generic real number generates translations in the coordinate space: for an infinitesimal $a$ we have
 \be \hat x e^{\hat p a} | x\rangle = \hat x (1+\hat pa) | x\rangle = i(x+a) e^{\hat p a}  |x\rangle, \ee
 where, in the second step, we have used the canonical commutators in (\ref{can-comm}). This means \be e^{\hat p a} | x\rangle =   |x+a\rangle \ee
  (a possible overall factor $k(a,x)$ in front of $|x+a\rangle$ can be set to one by a suitable definition of $\hat p$).
 From here we can construct the entire spectrum of $\hat p$. By applying $e^{\hat p a}$ on $\int dx |x\rangle$ one discovers that this is an eigenstate with zero momentum, and by applying $e^{-p\hat x}$ on it, where $p$ is a generic real number, one generates all possible eigenstates $|p\rangle$:
 \be |p\rangle = \frac1{\sqrt{2\pi}}\int dx \, e^{-p\hat{x}} |x\rangle   =  \frac1{\sqrt{2\pi}}\int dx \, e^{-ipx} |x\rangle,  \quad \iff  \quad  \langle x|p\rangle = \frac1{\sqrt{2\pi}} e^{i px} \label{fromxtop}\ee
 where the factor $1/\sqrt{2\pi}$ has been introduced to ensure the normalization condition 
 \be \langle p'|p\rangle = \delta(p+p'),\ee
 which, again, leads to the completeness relation $\int |p\rangle \langle p|\eta = 1$. 
 The states $|p\rangle$  satisfy
 \be \hat p  |p\rangle =ip  |p\rangle. \ee
 There are no other eigenstates as $i \hat p$ is self-adjoint with respect to the positively defined metric $\langle .|.\rangle_\eta \equiv \langle .|\eta|.\rangle$ and, therefore, $\hat p$ can only have purely imaginary eigenvalues. 
 
 The Dirac-Pauli quantization may look strange at first sight, but it can be seen as a complex canonical transformation performed on variables quantized in the ordinary way: $x\to ix$, $p\to -i p$.
 
 \begin{table}[t]
\centering
\begin{tabular}{ |p{3.59cm}|p{2.4cm}|p{2.4cm}|p{3.5cm}|p{3.5cm}| }
\hline
\rowcolor{red!80!green!40!yellow!10}Canonical variable& $\hat x$ on states & $\hat p$ on states & $\hat x$ on functions
 & $\hat p$ on 
functions \\
\hline
\vspace{0.2cm}
Dirac-Pauli variable \vspace{0.cm} & \vspace{-1.15cm} \bea \hat x |x\rangle \hspace{-0.3cm}&=& \hspace{-0.3cm} i x |x\rangle \nonumber  \\ \hat x |p\rangle  \hspace{-0.3cm}&=& \hspace{-0.3cm} -\frac{d}{dp} |p\rangle \nonumber \eea  \vspace{-0.5cm} & \vspace{-1.15cm} \bea \hat p |p\rangle  \hspace{-0.3cm}&=& \hspace{-0.3cm} i p |p\rangle \nonumber  \\  \hat p |x\rangle  \hspace{-0.3cm}&=& \hspace{-0.3cm} \frac{d}{dx} |x\rangle \nonumber \eea  \vspace{-0.5cm} &  \vspace{-1.15cm} \bea \langle x|\hat x|\psi\rangle  \hspace{-0.3cm}&=& \hspace{-0.3cm} -i x\langle x|\psi\rangle \nonumber  \\  \langle p|\hat x |\psi\rangle  \hspace{-0.3cm}&=& \hspace{-0.3cm} -\frac{d}{dp} \langle p|\psi\rangle \nonumber \eea  \vspace{-0.5cm}
 &  \vspace{-1.15cm} \bea \langle x|\hat p|\psi\rangle  \hspace{-0.3cm}&=& \hspace{-0.3cm} \frac{d}{dx}\langle x|\psi\rangle \nonumber  \\  \langle p|\hat p |\psi\rangle  \hspace{-0.3cm}&=& \hspace{-0.3cm} -ip \langle p|\psi\rangle \nonumber \eea  \vspace{-0.6cm}  \\
\hline
\vspace{0.2cm}
Ordinary variable \vspace{0.cm} & \vspace{-1.15cm} \bea \hat x |x\rangle \hspace{-0.3cm}&=& \hspace{-0.3cm}  x |x\rangle \nonumber  \\ \hat x |p\rangle  \hspace{-0.3cm}&=& \hspace{-0.3cm} - i \frac{d}{dp} |p\rangle \nonumber \eea  \vspace{-0.5cm} & \vspace{-1.15cm} \bea \hat p |p\rangle  \hspace{-0.3cm}&=& \hspace{-0.3cm}  p |p\rangle \nonumber  \\  \hat p |x\rangle  \hspace{-0.3cm}&=& \hspace{-0.3cm} i\frac{d}{dx} |x\rangle \nonumber \eea  \vspace{-0.5cm} &  \vspace{-1.15cm} \bea \langle x|\hat x|\psi\rangle  \hspace{-0.3cm}&=& \hspace{-0.3cm}  x\langle x|\psi\rangle \nonumber  \\  \langle p|\hat x |\psi\rangle  \hspace{-0.3cm}&=& \hspace{-0.3cm} i\frac{d}{dp} \langle p|\psi\rangle \nonumber \eea  \vspace{-0.5cm}
 &  \vspace{-1.15cm} \bea \langle x|\hat p|\psi\rangle  \hspace{-0.3cm}&=& \hspace{-0.3cm} -i\frac{d}{dx}\langle x|\psi\rangle \nonumber  \\  \langle p|\hat p |\psi\rangle  \hspace{-0.3cm}&=& \hspace{-0.3cm} p \langle p|\psi\rangle \nonumber \eea  \vspace{-0.6cm}  \\
\hline
\end{tabular}
\caption{\it Basic properties of a Dirac-Pauli variable (and its conjugate momentum) compared to the ordinary case. These properties are derived in the text or are simple extensions of the properties derived in the text. }
\label{table:DP}
\end{table}

 In Table \ref{table:DP} the basic properties of a Dirac-Pauli variable are summarized.

 \subsubsection{Making the wave functions normalizable}
 
 Let us now come back to our original problem, the non-normalizability of the wave functions. For the sake of definiteness, we  consider again the Pais-Uhlenbeck model with $\omega_1\neq \omega_2$
 and assume that $q_2$ is a Dirac-Pauli variable, while $q_1$ is an ordinary one. Then we obtain
 \beq\label{eq:psi0div2}
 \psi_{ 0}(q_1,q_2) \propto \exp\bigg(\frac{-q_1^2 \omega_1 \omega_2 -   q_2^2}{2}  (\omega_1 + \omega_2)  + q_1 q_2\omega_1 \omega_2\bigg),\eeq
 which is now normalizable:
 \be \langle 0|0\rangle=\int dq_1 dq_2 \langle 0|q_1, -q_2\rangle \langle q_1, q_2| 0\rangle   = \int dq_1dq_2 \psi_{ 0}(q_1,-q_2)^*\psi_{ 0}(q_1,q_2) <\infty,\ee
 where we have used the decomposition of the identity in terms of eigenstates of the coordinate operators and we have taken into account Eq.~(\ref{identityDP}) for the Dirac-Pauli variable $q_2$.
Moreover, recall that we have required before $\langle 0|0\rangle$ to be positive; we fix  $\langle 0|0\rangle=1$ by appropriately choosing the normalization constant.   Then, by using (\ref{gen-comm}),  one can easily show that the state $|n_1,n_2\rangle$, where $n_{1,2}$ are the occupation numbers of $\tilde a_{1,2}$, has norm $(-1)^{n_1}$. So, not only the ground state, but all excited states are normalizable with this quantization. 

 At this point it is good to mention that Hawking and Hertog~\cite{Hawking:2001yt} proposed a way to deal with four-derivative degrees of freedom, but they ended up with non-normalizable wave functions. They then suggested solving the problem by integrating out $\dot q$. As we have seen, this issue does not arise if the appropriate quantization described above is performed (treating $q$ as an ordinary variable and $\dot q$ as a Dirac-Pauli one)
   
   Other consistent quantizations are possible~\cite{Bender:2007wu,Bender:2008gh}. For example, one could quantize $\tilde q_1$ \`a la Dirac-Pauli, treating instead $\tilde q_2$ as an ordinary variable (the  variables with a tilde have been defined in Eq.~(\ref{eq:canqp'})). We will address this point after having introduced the path-integral formulation  of the theory.
   
   A Dirac-Pauli quantization for the ghost of QG has not been studied yet and is a very interesting topic for future research. By analogy with the results obtained in the Pais-Uhlenbeck model, one expects normalizable wave functions in the QG case too.

\subsubsection{Path-integral formulation}\label{Path-integral formulation}

 We now present the path-integral formulation of a theory with an arbitrary number of ordinary canonical variables $q_1, ..., q_n$ and Dirac-Pauli variables $\bar q_1, ..., \bar q_m$~\cite{Boulware:1983vw,Salvio:2015gsi}. A state with definite canonical coordinates is denoted here with 
 \be 
 |q\rangle = |q_1, ..., q_n, \bar q_1, ..., \bar q_m \rangle .\ee
 We are interested in understanding whether the quantization presented above is consistent in the presence of interactions. Even in ordinary quantum theories the real-time path integral is only a formal object, whose consistency at the rigorous level is unclear. For this reason, we consider  the imaginary-time path integral (what would be called the Euclidean path integral in a QFT). 
 
 In formulating a  quantum theory with the path integral one notices that the full information on the dynamics of the system is encoded in the object $\langle q_f| \exp{(-i H t)}|q_i \rangle$, where $|q_i \rangle$ and $|q_f \rangle$ are generic states with definite coordinates. Indeed, once this object is known we can determine how the wave function evolves in time. In the presence of some Dirac-Pauli variables one can do something similar, but one inserts an operator $\eta$ defined by
  \be \eta |q_1, ..., q_n, \bar q_1, ..., \bar q_m \rangle\equiv |q_1, ..., q_n, -\bar q_1, ..., -\bar q_m \rangle . \label{eta-general}\ee
Namely, instead of considering $\langle q_f| \exp{(-i H t)}|q_i \rangle$, one tries to evaluate $\langle q_f| \eta \exp{(-i H t)}|q_i \rangle$. This is convenient for reasons that will become apparent soon, but note that $\langle q_f| \eta \exp{(-i H t)}|q_i \rangle$ encodes the full dynamical information just like $\langle q_f|\exp{(-i H t)}|q_i \rangle$ as they both give the matrix elements of the time-evolution operators with respect to a complete basis.  

Working with an imaginary time $t\to - i \tau$, one is thus interested in computing the matrix element $\langle q_f| \eta \exp{(-H \Delta \tau)}|q_i \rangle$, where $\Delta \tau$ is some imaginary-time interval. This, as usual, can be done by decomposing $\Delta \tau$ in the sum of a very large number  $N$ of very small intervals $d\tau$, i.e. $d\tau \equiv \Delta \tau/N$. By writing $\exp{(-H \Delta \tau)} = \Pi_{j=1}^N \exp{(-H d\tau)}$ and inserting $N-1$ times the identity $\int dq |q\rangle \langle q|\eta = 1$ one ends up with
\be \langle q_f| \eta \,e^{-H \Delta \tau}|q_i \rangle  = \int \prod_{j=1}^{N} \langle q_j|\eta\, e^{- H d\tau} |q_{j-1} \rangle \prod_{k=1}^{N-1} d q_k, \ee
where $q_N\equiv q_f$ and $q_1\equiv q_0$.
To evaluate $\langle q_j|\eta \exp{(- H d\tau)} |q_{j-1} \rangle$ we insert the identity in the form $\int dp_{j-1} \eta|p_{j-1}\rangle\langle p_{j-1}|=1$:
\be \langle q_j|\eta \exp{(- H d\tau)} |q_{j-1} \rangle = \int dp_{j-1} \langle q_j|p_{j-1}\rangle \langle p_{j-1} |e^{-Hd\tau} |q_{j-1} \rangle = \int \frac{dp_{j-1}}{2\pi} e^{ip_{j-1}(q_j-q_{j-1}) - \bar H(q_{j-1}, p_{j-1}) d\tau}, \ee 
where we have used Eq.~(\ref{fromxtop}) and defined 
\be \bar H(q,p) \equiv \frac{\langle p| H | q\rangle}{ \langle p | q\rangle}.\ee 
Here we use a compact notation where the indices and sums over the various degrees of $q_1, ... , q_n$ and $\bar q_1, ...., \bar q_m$ are understood.
By letting $N\to \infty$ one thus obtains the imaginary-time path integral
\be   \boxed{\langle q_f| \eta \,e^{-H \Delta \tau}|q_i \rangle  = \int \delta q \delta p \, e^{ \int d\tau (i pq'-\bar H(q,p))}} \qquad \mbox{where} \quad  \delta q \delta p = \frac{dp_0}{2\pi}\lim_{N\to \infty}  \prod_{j=1}^{N-1} \frac{dq_j dp_j}{2\pi} ,  \label{PIEuclidean}\ee 
a prime denotes a derivative w.r.t.~$\tau$, the integral over $\tau$ is from an initial time $\tau_i$ and a final time $\tau_f$ such that $\Delta\tau = \tau_f-\tau_i$  and it is understood that the integral over $\delta q$ is performed only over those configurations that satisfy $q(\tau_i) = q_i$ and $q(\tau_f) = q_f$.

We see that, modulo the usual subtleties related to the integration over an infinite-dimensional functional space that are present in any quantum theory, the only requirement for the existence of the path integral is that the real part of  $\bar H(q, p)$ (not\footnote{In ordinary quantum theories $\bar H(q, p) = H(q, p)$, but in the presence of Dirac-Pauli variables this is not generically the case because of the extra $i$ appearing in the eigenvalues of the Dirac-Pauli coordinates and momenta.} the classical Hamiltonian $H(q,p)$) be bounded from below and that $\bar H(q, p)$ diverge fast enough when the canonical coordinates tend to infinity (so that the integrations over $q$ and $p$ converge). 

These conditions are satisfied in the Pais-Uhlenbeck model where $q_1$ is quantized in the ordinary way and $q_2$ is quantized \`a la Dirac-Pauli, at least when the interaction term $V$ is bounded from below\footnote{If one introduces a more complicated interaction that depends on the other coordinate and momenta $V(q,p)$, the condition is that Re$\bar V(q,p)$ be bounded from below.} (the usual condition). Indeed, from the Hamiltonian (\ref{eq:HOstro}) it follows
\be \bar H(q, p) =  i  p_1 q_2 + \frac{1}{2}p_2^2 + \frac{\omega_1^2+\omega_2^2}{2} q_2^2 +\frac{\omega_1^2\omega_2^2}{2}q_1^2+V(q_1), \label{PUUbar} \ee 
which has the required properties. 
For the Pais-Uhlenbeck model  the Euclidean  path integral is
\beq\langle q_f| \eta \,e^{-H \Delta \tau}|q_i \rangle  = 
\int \delta q_1\delta q_2\delta p_1 \delta p_2~\exp\bigg[\int d\tau  (i p_1 q'_1 + i p_2 q'_2-\bar H(q, p)) \bigg].\eeq
This expression can be further simplified since some integrations can be explicitly performed. Given the first term in (\ref{PUUbar}), the $\delta p_1$ integral gives $\delta(q_2-q'_1)$, such that the $\delta q_2$ path integral just fixes $q_2 = q'_1$.
Next, the remaining terms in $\bar H$ are a sum of positive squares and $V(q_1)$ so all other integrals are convergent assuming that $V$ is bounded from below.
Performing the remaining integrals,  one finds the Lagrangian Euclidean path integral:
\beq\langle q_f| \eta \,e^{-H \Delta \tau}|q_i \rangle  \propto \int \delta q \,  \exp\bigg[- \int d\tau  L_E(q) \bigg], \label{PathPU}\eeq
where the classical Euclidean Lagrangian is
\beq L_E =\frac12 \bigg(\frac{d^2q}{d\tau ^2}\bigg)^2
 + \frac{\omega_1^2+\omega_2^2}{2}
 \bigg(\frac{dq}{d\tau}\bigg)^2+
 \frac{ \omega_1^2 \omega_2^2}{2} q^2 + V(q) \label{LEPU}
.\eeq
The Lagrangian path integral appears to be well-defined as $L_E$ is bounded from below.

The expression in~(\ref{LEPU}) also allows us to study the classical limit. Going back to real time one obtains precisely the Lagrangian we started from, Eq.~(\ref{eq:LagO}). 
 As discussed in Sec.~\ref{The Pais-Uhlenbeck model},  for some interactions $V(q)$ (bounded from below and above) there are stable solutions. In a generic theory, one expects that the requirement of having stable solutions put stringent conditions on the possible interactions, which so far have not been fully classified. 
 The path integral formulation tells us that, in the classical limit, the dynamics is dominated by the solution(s) with least Euclidean action. In the Pais-Uhlenbeck case these correspond to time-independent solutions that minimize the full potential $
 \frac{ \omega_1^2 \omega_2^2}{2} q^2 + V(q)$.
 All unbounded solutions, if any, should be negligible in the classical limit as the derivative terms always contribute positively to the Lagrangian in (\ref{LEPU}). As usual, perturbations around a given solution should be computed through the path integral and, given that the path integral appears to be well-defined no pathologies are expected. Therefore, it is possible that the Dirac-Pauli quantization could solve the potential problems raised by the Ostrogradsky theorem.

 The path integral (\ref{PathPU}) makes it clear that, if $V(q)$ is chosen to be non-negative everywhere, no negative energies can be present: if they did we should observe a divergence of $\langle q_f| \eta \exp{(-H \Delta \tau)}|q_i \rangle$ as $\Delta \tau \to \infty$, but the right-hand side of (\ref{PathPU}) does not diverge in that limit as the Lagrangian is a sum of positive terms.
 
 Another issue is that in a theory where the Hamiltonian $H$ is self-adjoint with respect to an indefinite norm (and nothing else is known) there is no theorem guaranteeing the reality of the energy spectrum. However, it is still possible that the spectrum is real, as we have seen in the case of the unequal-frequency Pais-Uhlenbeck model in Sec.~\ref{The Pais-Uhlenbeck model: quantum}. Even if one introduces a non-trivial interaction term $V\neq 0$ in the Pais-Uhlenbeck model with generic unequal frequencies, no complex energies can appear as long as $V$ is small enough that perturbation theory can be trusted: indeed, a complex energy would require a zero-norm state, but only positive and negative norm eigenstates of $H$ with no degeneracies are found in Sec.~\ref{The Pais-Uhlenbeck model: quantum}.    In a theory where some of the eigenvalues of $H$ turn out to be complex one should find a sensible interpretation for them. A possible interpretation could be that those states are unstable and some of them (the ones with eigenvalues with positive imaginary parts) lead to a violation of causality\footnote{Nevertheless the commutators between any two field operators at points separated by a spacelike distance are zero~\cite{Lee:1970iw}, like in usual QFT. In QG this property can be easily proved by using the expansion of the free ghost field in creation and annihilation operators introduced as in Sec.~\ref{The Pais-Uhlenbeck model: quantum} and then by applying the unitary operator that transforms the free ghost field in the interacting one.}~\cite{ColemanAcausality,Grinstein:2008bg}. However, in Ref.~\cite{Salvio:2017xul} it was pointed out that there are some conditions  to be fulfilled in order for this violation of causality to be observable and it is easy to engineer a model where these conditions are not met.

Let us come back to the path integral. What would have happened if we had used a different quantization? One could have quantized $\tilde q_1$ \`a la Dirac-Pauli and $\tilde q_2$ as an ordinary variable (the  variables with a tilde have been defined in Eq.~(\ref{eq:canqp'}) when $\omega_1 \neq \omega_2$). Then, one would have obtained
\beq\langle \tilde q_f| \eta \,e^{-H \Delta \tau}|\tilde q_i \rangle  = 
\int \delta \tilde q_1\delta \tilde q_2\delta \tilde p_1 \delta \tilde p_2~\exp\bigg[\int d\tau  (i \tilde p_1 \tilde q'_1 + i \tilde p_2 \tilde q'_2-\bar H(\tilde q, \tilde p)) \bigg],\eeq
where 
\beq \bar H(\tilde q, \tilde p)= \frac12 (\tilde p^{ 2}_1  +\omega_1^2 \tilde q_1^{ 2}) + \frac12 (\tilde p^{2}_2 + \omega_2^2 \tilde q^{ 2}_2) + \bar V(\tilde q_2, \tilde p_1) \label{DecH}\eeq
and, according to Eq.~(\ref{eq:canqp}),
 \be  \bar V(\tilde q_2, \tilde p_1) =V(\frac{ \tilde{q}_2 -  i \tilde{p}_1/\omega_1}{\sqrt{\omega_1^2 - \omega_2^2}}).
  \ee
  Given that $V$ is computed in the complex quantity  $(\tilde{q}_2 -  i \tilde{p}_1/\omega_1)/\sqrt{\omega_1^2 - \omega_2^2}$, the requirement that Re$\bar H(\tilde q, \tilde p)$ is bounded from below leads to very peculiar conditions on the function $V$, which seems very hard to be fulfilled for reasonable $V$ and thus very hard to be kept in generalizing these results to QG. Therefore, while other quantizations could still be consistent, dedicated studies of these alternative path-integral quantizations in the presence of interactions are not known.

 The computation of the Lagrangian path integral has been carried out here within the Pais-Uhlenbeck model. We have used explicitly that some variables are quantized \`a la Dirac-Pauli. If  a Dirac-Pauli quantization for QG will be provided one could also perform the same calculation in QG. One expects that the Lagrangian path-integral for QG is consistent if the classical Euclidean Lagrangian is bounded from below, 
 which is the case for some choices of the parameters, but there is no substitute of a complete calculation to reach this conclusion. Such calculation would also provide a non-perturbative definition of quantum QG.

 \subsubsection{Probabilities}\label{Probabilities}
 
 We now turn to the possible definitions of probabilities in the presence of ghosts. We have learned in Secs~\ref{Renormalizability} and~\ref{The Pais-Uhlenbeck model: quantum} that both the renormalizability of QG and the requirement that the quantum Hamiltonian must be bounded from below lead to the presence of an indefinite metric. This raises problems in defining the probability that a certain event occurs. In quantum mechanics, the possible outcomes of the measurement of an observable $A$  (a self-adjoint operator, $A^\dagger = A$) are in one-to-one correspondence with the eigenstates $|a\rangle$ of $A$ with probabilities given by the Born rule
 \be P(\psi \to a) = \frac{|\langle a|\psi \rangle|^2}{\langle a|a\rangle \langle \psi|\psi\rangle}, \label{Obr}
 \ee
 where $|\psi\rangle$ is the state of the system before the measurement. If some of the states have negative norms, the direct application of the Born rule in the presence of ghosts leads to some negative probabilities.
 
 Since $P(\psi \to a)$ can be negative only when the denominator $\langle a|a\rangle \langle \psi|\psi\rangle$ is negative a first idea could be to substitute (\ref{Obr}) with the following modified Born rule:
  \be P(\psi \to a) = \frac{|\langle a|\psi \rangle|^2}{|\langle a|a\rangle \langle \psi|\psi\rangle|}, \label{Obrp}
 \ee
 However, (\ref{Obrp}) does not generically satisfy another basic requirement, that the sum of $P(\psi \to a)$ over all possible eigenvalues $a$ is 1.  This is because 
\be  \sum_a \frac{|\langle a|\psi \rangle|^2}{|\langle a|a\rangle \langle \psi|\psi\rangle|}= \sum_a \frac{\langle \psi|a \rangle\langle a|\psi \rangle}{|\langle a|a\rangle \langle \psi|\psi\rangle|} \ee
and here generically we have
\be \sum_a \frac{|a\rangle \langle a|}{\left|\langle a|a\rangle \right|}\neq 1. \ee
Indeed, if we assume the eigenstates $|a\rangle$ to form a complete basis and decompose an arbitrary  state $|\alpha\rangle$ as $|\alpha\rangle=\sum_{a'} \alpha_{a'} |a'\rangle$, where $\alpha_{a'}$ are complex numbers,  we have
\be  \sum_a \frac{|a\rangle \langle a|\psi\rangle}{\left|\langle a|a\rangle \right|} = \sum_{aa'}\frac{\alpha_{a'}  |a\rangle \langle a|a'\rangle}{\left|\langle a|a\rangle \right|} \label{action}\ee
and in general $\langle a|a'\rangle/\left|\langle a|a\rangle \right|$ is not equal to $\delta_{aa'}$ because some of the states can have negative norm. This is what some people call the ``unitarity problem"  (we do not use this terminology here as the time evolution operator {\it is} unitary w.r.t. indefinite norm).

 We now discuss the most popular ways to address this problem. 
 \subsubsection*{\it Lee-Wick idea}

Lee and Wick~\cite{Lee:1969fy} proposed that a theory with an indefinite metric can still have a unitary $S$-matrix provided that all stable states have positive norm.
 Since the $S$-matrix connects only asymptotic states that, by definition, are stable, one expects that under this hypothesis the transition probabilities between asymptotic states are positive and add up to one.
 The Lee-Wick idea has been studied in the context of QG in a number of papers~\cite{Tomboulis:1980bs,Antoniadis:1986tu,Hasslacher:1980hd,Salvio:2016vxi,Anselmi:2017yux,Anselmi:2017lia,Anselmi:2017ygm,Anselmi:2018kgz,Anselmi:2018ibi}.

 To understand more in detail this idea, let us denote with $|\sigma\rangle$ and $|\sigma'\rangle$ two generic stable states and consider the $S$-matrix elements 
 \be S_{\sigma'\sigma}\equiv  
 \langle \sigma'|S|\sigma\rangle,
 \label{Smatrix}
 \ee
 where we have normalized  $|\sigma\rangle$ and $|\sigma'\rangle$ to 1 (the Lee-Wick hypothesis implies that the norm of stable states are positive and therefore can be normalized to 1). The operator $S\equiv \lim_{\Delta t \to \infty} U(\Delta t)$ is unitary with respect to the indefinite norm by construction, but we are interested in proving the unitarity of the $S$-matrix in~(\ref{Smatrix}) because this is what would allow us to claim that the probabilities add up to one: indeed, using the standard Born rule (\ref{Obr}) leads to 
 \be \sum_{\sigma'} P(\sigma \to  \sigma') =\sum_{\sigma'} 
 |\langle \sigma'|S|\sigma\rangle|^2
  = \sum_{\sigma'}  S_{\sigma'\sigma}^*  S_{\sigma'\sigma}.  \ee 
  Now, one can rewrite
 \be \sum_{\sigma'} 
 |\langle \sigma'|S|\sigma\rangle|^2
  = \sum_{\sigma'}
 \langle \sigma|S^\dagger|\sigma'\rangle \langle \sigma'|S|\sigma\rangle
  \ee
 and this expression would be equal to  1 in two cases
 \begin{enumerate}
 \item if $\sum_{\sigma'}|\sigma'\rangle \langle \sigma'|=1$ or, more generally,
 \item  if $S|\sigma\rangle$ can be written as a linear combination of the stable states only.
 \end{enumerate}
  The first condition cannot be true because we know there are negative norm states,  which can never be written as linear combinations of positive-norm states only; 
 indeed, in the presence of negative norm states $\sum_{\sigma'}|\sigma'\rangle \langle \sigma'|=1$ is replaced by 
 \be \sum_{\sigma'}|\sigma'\rangle \langle \sigma'|=1 - \Pi_- \label{Pim},\ee where $\Pi_-$ is the projector on the negative-norm subspace.
  So one has to assume  Condition 2, which, although plausible (as one expects $S$ to connect stable states with stable states only), {\it has to be proved}. 
 To see when the important probabilistic condition $\sum_{\sigma'} 
 |\langle \sigma'|S|\sigma\rangle|^2 =1$ is satisfied it is convenient to rewrite it in a form that can be more easily verified by an explicit calculation. To do so we note that
 \be 
 \sum_{\sigma'}  
 \langle \sigma|S^\dagger |\sigma'\rangle  \langle \sigma' |S|  \sigma\rangle = 1 -\langle\sigma|S^\dagger \Pi_-S|  \sigma\rangle, \ee
 where we have used Eq.~(\ref{Pim}). By writing as usual $S \equiv 1+i T$ one has 
 \be \langle\sigma|S^\dagger \Pi_-S|  \sigma\rangle = \langle\sigma|T^\dagger \Pi_-T|  \sigma\rangle, \ee
 which follows from $\Pi_- |\sigma\rangle = 0$. The unitarity of $S$  implies $i(T^\dagger-T) = T^\dagger T$ and, by taking the diagonal matrix element $T_{\sigma\sigma} \equiv \langle \sigma | T | \sigma \rangle $ and using once again Eq.~(\ref{Pim}), 
 \be 2 {\rm Im} T_{\sigma \sigma} =  \sum_{\sigma'}  \langle \sigma|T^\dagger |\sigma'\rangle  \langle \sigma' |T|  \sigma\rangle +  \langle \sigma|T^\dagger \Pi_-T|  \sigma\rangle\ee
 Given that $\Pi_-$ can be written as $\sum_{g}   |g\rangle  \langle g |$  where $|g\rangle$ represents a complete basis on the negative-norm subspace, we see that the condition that the probabilities sum up to one is equivalent to the condition that the ghost states $|g\rangle$ do not contribute to the imaginary part of the  forward scattering amplitude, represented here by $T_{\sigma\sigma}$.
 Ref.~\cite{Anselmi:2017ygm} has recently found that this condition is satisfied if one modifies appropriately the prescription to determine the ghost propagator\footnote{See also~\cite{Donoghue:2018izj,Abe:2017abx} for other discussions about unitarity.}.

One  issue is  that, in order to claim that the negative norm states are unstable, 
which is a basic assumption of the Lee-Wick proposal, one  needs a consistent way of computing the probability of ghost decays;
otherwise how do we tell if the ghost is unstable or not? Since there is  one ghost field in QG the  use of the standard Born rule (\ref{Obr}) to compute this probability leads to a negative number. This is not necessarily a non-sense as Lee and Wick proposed to consider as physical states only the asymptotic ones and regard the ghost just as a virtual state, which is not directly observable. In this case it might be consistent to assign negative probabilities to such somewhat unobservable events, as pointed out by Feynman~\cite{Feynman}.

However, one can also argue that the Lee-Wick proposal might not address all potential problems because scattering theory (described by the $S$-matrix) is not the only application of quantum mechanics. 
  
\subsubsection*{\it  Defining Positive norms }

 Although renormalizability and the existence of a state of minimum energy lead to an indefinite metric, one can still try to define positively defined metrics with the desired property: positive probabilities that add up to one when used in the Born rule. This possibility was studied in a number of articles~\cite{Bender:2002vv,Bender:2007nj,Bender:2007wu,Salvio:2015gsi,Mannheim:2015hto,Mannheim:2017apd,Strumia:2017dvt, Raidal:2016wop}.

Let us consider an example of a positively defined metric. The path-integral formula (\ref{PIEuclidean}) suggests to consider the $\eta$-metric  $\langle . | .\rangle_\eta \equiv \langle . |\eta|.\rangle$, where $\eta$ is defined for a generic theory in Eq.~(\ref{eta-general}). This metric is  positively defined because 
\be \langle q'_1, ..., q'_n, \bar q'_1, ..., \bar q'_m|\eta|q_1, ..., q_n, \bar q_1, ..., \bar q_m \rangle = \prod_{j=1}^n\delta(q_j - q_j') \prod_{k=1}^m\delta(\bar q_j - \bar q_j')  \label{eta-norm-gen} \ee
and $|q_1, ..., q_n, \bar q_1, ..., \bar q_m \rangle$ is complete.
In (\ref{eta-norm-gen}) we used (\ref{xx'inner}) for the Dirac-Pauli variables $\bar q_1, ... ,\bar q_m$ and the usual normalization $\langle q_j|q_j'\rangle = \delta(q_j - q_j')$ for the ordinary variables $q_1, ... , q_n$. 
 The $\eta$-metric can be used to compute the probabilities of measuring $q_1, ..., q_n, \bar q_1, ..., \bar q_m$ and the corresponding conjugate momenta (in the case of Dirac-Pauli variables, the outcomes of an experiment can be identified with the imaginary parts of the eigenvalues). Below we will show that the probabilities add up to one.
 
 Before doing so we generalize this approach to other observables. First, we have to clarify the meaning of ``observables" in this context.
  An observable $A$ is represented by an operator with a complete set of eigenstates, $|a\rangle$.  
 Indeed, in this case we can define a positively defined metric in the following way. Let us {\it define} an operator $P_A$ through\footnote{This defines $P_A$ because an operator is defined once we give all matrix elements in a complete basis.}
\be \langle a'| P_A|a\rangle \equiv \delta_{aa'}. \label{defPA}\ee
 Note that $P_A$ satisfies $P_A^\dagger=P_A$
 and depends in general on $A$. The new positively defined metric is defined by 
 \be  \langle\psi_2|\psi_1\rangle_{A}\equiv \langle \psi_2|P_A|\psi_1\rangle,  \ee
 where $|\psi_{1,2}\rangle$ are generic states. By using this new metric one can define the probabilities with the usual Born rule: the probability that the outcome of an experiment will measure $a$ for an observable $A$ given that the state before the measurement is $|\psi\rangle$ is given by
 \be P(\psi\rightarrow a)\equiv \frac{\left|\langle a| \psi\rangle_A \right|^2}{\langle a|a\rangle_A \langle \psi|\psi\rangle_A}.\ee
 These probabilities indeed satisfy the basic properties: they are positive and they add up to one:
 \be \sum_a P(\psi\rightarrow a) = \sum_a 
 \frac{\langle \psi | a \rangle_A \langle a | \psi\rangle_A}{\langle a|a\rangle_A \langle \psi|\psi\rangle_A} = \frac{\langle \psi | P_A}{\sqrt{ \langle \psi|\psi\rangle_A}} \left(\sum_a \frac{|a\rangle\langle a| P_A}{\langle a|P_A| a\rangle}\right)\frac{|\psi\rangle}{\sqrt{ \langle \psi|\psi\rangle_A}} = 1,
 \ee
where we used
\be \sum_a \frac{|a\rangle\langle a| P_A}{\langle a|P_A| a\rangle} =1, \ee 
which follows from the completeness of $\{|a\rangle\}$ and the defining property of $P_A$, eq.~(\ref{defPA}). Note that this result also holds for time-dependent $|\psi\rangle$ and, therefore, probability is conserved under time evolution.
In the specific case when $\langle a|a\rangle$ is either positive or negative (it never vanishes) an explicit expression for $P_A$ is (after having normalized the state in a way that $\langle a|a\rangle = \pm 1$)
\be P_A\equiv \Pi^A_+-\Pi^A_-, \label{new-metric}\ee
where $\Pi^A_+$ and $\Pi^A_-$ are the projectors on the positive norm and negative norm eigenstates of $A$, respectively.

 \subsection{Cosmology}\label{Cosmology}
 
 In practice the cosmological predictions of QG would be basically those of a standard QFT coupled to Einstein gravity if it were not for the $W^2$ term. This term, as we have seen, corresponds to a spin-2 ghost with mass $M_2 = f_2 \bp/\sqrt{2}$. Therefore, unless one takes $f_2$ really tiny, the only significant effects of the ghost occur in an inflationary context. We will focus then on the inflationary behaviour of the theory here. 

 The first step in studying the cosmological applications of the theory is to find an FRW metric that satisfies the classical equations. 
From the experience gained with the Pais-Uhlenbeck model in Sec.~\ref{Path-integral formulation}, one expects that the classical limit provides precisely the classical action we started from, Eqs.~(\ref{Lgravity3}),~(\ref{Lmatter}) and~(\ref{Lnonminimal}). This is what is assumed basically in   the entire literature on the subject. The actual proof of this property would be a significant progress in the understanding of QG. 
 
The FRW metric is 
 \be ds^2 =  dt^2 -a(t)^2\delta_{ij}dx^idx^j, \label{FRW} \ee
 where $a$ is the scale factor
and 
 we have neglected  the spatial curvature parameter as during inflation the energy density is dominated by the scalar fields. 
 The metric in~(\ref{FRW}) leads to standard Friedmann equations as the $W^2$ term vanishes on conformally flat metrics and does not contribute to  the equations of motion. 
 When the hypothesis of homogeneity and isotropy is relaxed the $W^2$ term contributes instead and its effect has been studied  in a number of works~\cite{Berkin:1991nb,Hamada:2006ei,Clunan:2009er,Weinberg:2009wa,Nelson:2010wp,Deruelle:2010kf,Myung:2014jha,Shapiro:2014fsa,Myung:2015vya,Ivanov:2016hcm,Salvio:2017xul} (see Ref.~\cite{Salvio:2017xul} for a general treatment), where the perturbations around the FRW metrics were considered. We do not reproduce the calculations here as they are performed in detail in the original articles. One of the most important results obtained so far is that all perturbations found by solving the linear equations around the FRW metric remain bounded as time passes by~\cite{Salles:2014rua,Ivanov:2016hcm,Tokareva:2016ied,Salvio:2017xul,Salles:2017xsr}, contrarily to what one would naively expect from the Ostrogradsky theorem. Moreover, by quantizing these linear perturbations with an indefinite metric (with an appropriate generalization of Sec.~\ref{The Pais-Uhlenbeck model: quantum}) one obtains that the conserved Hamiltonian of the full system is bounded from below~\cite{Salvio:2017xul}. What happens beyond the linear order, however, has not been discussed in detail and is an important target for future research. 

In QG there are several possible inflaton candidates. First, QG gives a natural implementation of Starobinsky's inflationary model~\cite{Starobinsky:1980te} as the $R^2$ is mandatory in order to have renormalizability. Furthermore, other possible scalar fields can participate: at the very least the theory should contain the Higgs boson, which has been discovered at the Large Hadron Collider. A detailed analysis of the inflationary dynamics and observable predictions in some specific realizations of the QG scenario is provided in~\cite{Salvio:2014soa,Kannike:2015apa,Salvio:2017xul}.

\subsection{Black holes}

After the discovery of gravitational waves interpreted as the product of a binary black hole merger~\cite{Abbott:2016blz}, the interest in black hole solutions have increased. Therefore, it is important to study the existence and properties of static spherically symmetric solutions in QG, 
 where the metric is given in spherical coordinates $\{r,\theta,\varphi\}$ by two functions $f_1$ and $f_2$ of $r$:
 \be ds^2 =f_1(r)dt^2 - \frac{dr^2}{f_2(r)} - r^2(d\theta^2+\sin^2\theta d\varphi^2). \ee

 This have been initiated in a number of articles. The first work was  done by Stelle~\cite{Stelle:1977ry}, who computed the correction to Newton's law due to the extra gravitational terms. A first observation is that the Schwarzschild solution   of Einstein gravity in the vacuum 
 ($f_1(r) = f_2(r)$) is also a solution of the vacuum equations of QG (i.e. in the absence of matter)~\cite{Stelle:1977ry,Nelson:2010ig,Lu:2015cqa}.  Also, Refs.~\cite{Holdom:2002xy,Lu:2015cqa,Lu:2015psa,Cai:2015fia,Lin:2016kip,Goldstein:2017rxn,Lu:2017kzi,Kokkotas:2017zwt,Stelle:2017bdu} found numerically and studied new black hole solutions (not present in Einstein gravity) and Ref.~\cite{Holdom:2016nek} identified a new class of static spherically symmetric solutions without horizon (called the 2-2-hole), which can, nevertheless, mimic the Schwarzschild solution outside the horizon, with interesting implications for the black hole information paradox. 

Keeping in mind the Ostrogradsky theorem, an important question is whether a stable black hole (or pseudo black hole, such as the 2-2-hole) exists in the theory. Ref.~\cite{Lu:2017kzi} pointed out that the Schwarzschild solution is stable for large horizon radius $r_h$, but becomes unstable (see also~\cite{Myung:2013doa}) when $r_h$ is taken below a critical value set basically by the inverse ghost mass $\sim 1/M_2$ (see also~\cite{Stelle:2017bdu}); the endpoint of the instability is conjectured to be another black hole solution, which is not present in Einstein gravity and may be stable when $r_h$ is small. Ref.~\cite{Holdom:2016nek} considered the creation of a static spherically symmetric solution generated by a thin spherically symmetric shell of matter; 
 when the shell radius $l\lesssim r_h$ the new 2-2-hole is found. 

Once again in all these works the classical equations (valid as $\hbar\to 0$) of QG are taken to be those generated by the starting action in~(\ref{Lgravity3}), which is what we expect but, as pointed out in Sec.~\ref{Cosmology}, a proof is still missing in the literature.

\section{Reaching infinite energy}\label{Reaching infinite energy}

Given that QG (coupled to a general renormalizable matter sector) is renormalizable one can hope that the theory remains valid up to infinite energy. However, soon after the calculation of the gravitational $\beta$-functions of~\cite{Avramidi:1985ki} it was realized a major obstacle to UV-completeness: the $\beta$-function of $f_0^2$ in (\ref{RGGadim2}) is not negative for $f_0^2>0$ and, therefore,  the theory features a growth of $f_0$ as the energy increases, until perturbation theory in $f_0$ cannot be trusted anymore\footnote{Different statements in the literature (even recent)
appear because some results for the $\beta$ of $f_0$ (obtained before the correct results of~\cite{Avramidi:1985ki}) contained wrong signs.}. 

Then, a number of authors~\cite{Buchbinder:1989jd,Buchbinder:1992rb,Tomboulis:2015esa,Einhorn:2015lzy,Einhorn:2016mws,Einhorn:2017icw} explored the case $f_0^2<0$ claiming that asymptotic freedom can  be achieved for all couplings (both the gravitational and matter couplings) if the matter sector is chosen appropriately. Although such programme can lead to mathematically consistent asymptotically free theories, there is a big phenomenological problem when one chooses $f_0^2<0$. 

Let us consider for simplicity the case where the scalar $\zeta$ (corresponding to the $R^2$ term and introduced in Sec.~\ref{Einstein frame Lagrangian}) does not mix with other scalars (if any). Then, the squared mass of $\zeta$ equals $M_0^2 = f_0^2 \bp^2/2$ (see Table~\ref{table:dof}), which clearly indicates that for $f_0^2<0$ the scalar $\zeta$ is tachyonic. One way to obtain $M_0^2 = f_0^2 \bp^2/2$ is to use the Einstein frame Lagrangian in~(\ref{eq:LmatterE}) and~(\ref{Udef}) and compute its quadratic approximation for the small fluctuations around the flat spacetime. Another way is to calculate (directly in the Jordan frame) the propagator of $h_{\mu\nu} \equiv g_{\mu\nu} - \eta_{\mu\nu}$, a procedure that was originally performed in Ref.~\cite{Stelle:1976gc}, which obtained precisely the masses given in Table~\ref{table:dof}. This confirms that $f_0^2<0$ leads to a tachyonic instability\footnote{Similarly, $f_2^2<0$ leads to a tachyonic instability in the ghost sector  and, therefore, this case is commonly avoided as not even consistent with asymptotic freedom (for a discussion of the tachyonic case see, however, Refs.~\cite{Narain:2011gs,Narain:2016sgk,Narain:2017tvp}).}. Yet another way to see why $f_0^2<0$ is phenomenolgically problematic is to look at the Newtonian potential $V_N(r)$ due to the Lagrangian (\ref{Lgravity3})~\cite{Stelle:1977ry,Alvarez-Gaume:2015rwa}
\be V_N(r) = - \frac{G_NM}{r}\left(1 -\frac43 e^{-M_2r} +\frac13 e^{-M_0r} \right), \ee
where $G_N$ is Newton's gravitational constant and $M$ is the mass of the point particle generating the potential.
As noted even in the original article~\cite{Stelle:1977ry} by Stelle, this expression only gives an acceptable Newtonian limit for real $M_2$ and $M_0$ (i.e. for positive $f_2^2$ and $f_0^2$): otherwise one would obtain oscillating $1/r$ terms.

One could hope that a phenomenologically viable $f_0^2<0$ is achieved by introducing more scalars (besides $\zeta$). However, a general argument, which we now describe, indicates that this is not the case. Consider the Einstein frame potential $U$ (defined in Eq. (\ref{Udef})) along the $\zeta$-direction, which can be conveniently parameterised as 
\be U = \frac{1}{\zeta^4} \left[a_1+a_2 (\zeta^2-a_3)^2\right]\ee
where $a_1, a_3$ are suitable coefficients, which depend on the other scalar fields, while $a_2=3f_0^2\bp^4/8<0$ (having assumed $f_0^2<0$ here). A necessary condition for the existence of a minimum of $U$ is that 
\be \frac{\partial U}{\partial \zeta}  = 0, \qquad \mbox{that is} \qquad \zeta^2=\frac{a_1+a_2a_3^2}{a_2a_3}. \label{solz} \ee
Notice that, if the solution for $\zeta^2$ exists, that is $(a_1+a_2a_3^2)/a_2a_3 >0$, then it is unique. Moreover, note that $a_2<0$ implies that $U$ goes to a negative value as $\zeta\rightarrow \infty$. Therefore, there are only three possibilities
\begin{itemize}
\item There is no acceptable solution to (\ref{solz}) (no solution with $\zeta^2>0$).
\item The solution to (\ref{solz}) is a maximum  of the potential (or at most a saddle point once the other scalars are included).
\item The solution to (\ref{solz}) is a point of minimum of $U$, but occurs for a negative value of $U$ (in contradiction with the positive value of the observed cosmological constant).
 Indeed, if it corresponded to a positive value of $U$ then there would also  be a maximum (or a saddle point) given that $U$ goes to a negative value for $\zeta\rightarrow \infty$ and this would contradict the uniqueness of the solution in (\ref{solz}).
\end{itemize}

The conclusion is that  a minimum of $U$ (if any) must have $U<0$. This argument generalizes the situation illustrated in Fig.~\ref{Uomega}, where only the field $\zeta$ was considered.

\subsection{Conformal gravity as the infinite energy limit of quadratic gravity}

Given that the experiments lead us to take $f_0^2>0$, what happens when $f_0$ grows and leaves the domain of validity of perturbation theory? In Ref.~\cite{Salvio:2017qkx} (see also references therein), by using a perturbative expansion  in $1/f_0$, it was shown that, when $f_0$ grows  up to infinity in the infinite energy limit, the scalar due to the $R^2$ term  decouples from the rest of the theory and $f_0$ does not hit any Landau pole,
 provided that all scalars have asymptotically Weyl-invariant couplings (see below) and all other couplings approach fixed points. 
Then, QG can flow to a Weyl-invariant theory, a.k.a. conformal gravity, at infinite energy.  
Given the importance of Weyl invariance for the high-energy limit of QG, let us give some more details on this topic. A Weyl transformation acts as follows on the various fields (the metric $g_{\mu\nu}$, the scalars $\phi_a$, the fermions $\psi_i$ and the vectors $V_\mu^A$):
\be g_{\mu\nu}(x) \to  e^{2\sigma(x)} g_{\mu\nu}(x),\quad \phi_a(x)\rightarrow e^{-\sigma(x)}\phi_a(x), \quad \psi_i(x)\rightarrow e^{-3\sigma(x)/2}\psi_i(x), \qquad V^A_\mu\rightarrow V^A_\mu, \label{sigmaIntro}\ee
where $\sigma$ is a generic function of $x$.
A scalar has Weyl-invariant couplings when all dimensionful parameters vanish and $\xi_{ab}=-\delta_{ab}/6$. This precise value of $\xi_{ab}$ emerges because in this case the non-invariance of the kinetic term of the $\phi_a$ precisely cancels the non-invariance of the non-minimal couplings, Eq.~(\ref{Lnonminimal}).

The idea that one can approach a Weyl-invariant theory at large energy has been investigated in a number of articles~\cite{Fradkin:1978yf,Zee:1983mj,Shapiro:1994st,Hamada:2002cm,Hamada:2009hb,Donoghue:2016xnh,Donoghue:2018izj,Alvarez:2017spt}.
We do not reproduce the proof of Ref.~\cite{Salvio:2017qkx} because it is described in detail there, but some remarks are in order regarding the implications of this result.

It is important to note that the condition to have a UV fixed point guarantees  not only the UV-completeness of the QFT part\footnote{Some SM extensions including gauge fields, fermions and scalars can feature a UV fixed point for all couplings and their corresponding phenomenology have been studied~\cite{Giudice:2014tma,Holdom:2014hla,Pelaggi:2015kna,Mann:2017wzh,Pelaggi:2017abg}.} but also of the gravitational part of the theory  (when all parameters flow to their conformal value). This opens the road to the construction and study of relativistic field theories of {\it all} interactions that are fundamental, i.e. hold up to infinite energy. This scenario leads to several extra fields (in addition to those present in the SM) as the study of the one-loop $\beta$-functions of the SM reveals the presence of Landau poles. These new fields can then be used to explain in an innovative way the current pieces of evidence for physics beyond the SM (neutrino oscillations, dark matter, baryon asymmetry of the universe, etc.). This nearly unexplored field of research represents a very important target for future research.

\subsection{RGEs for conformal gravity and matter}

\begin{figure}[t]
\begin{center}
  \includegraphics[scale=0.7]{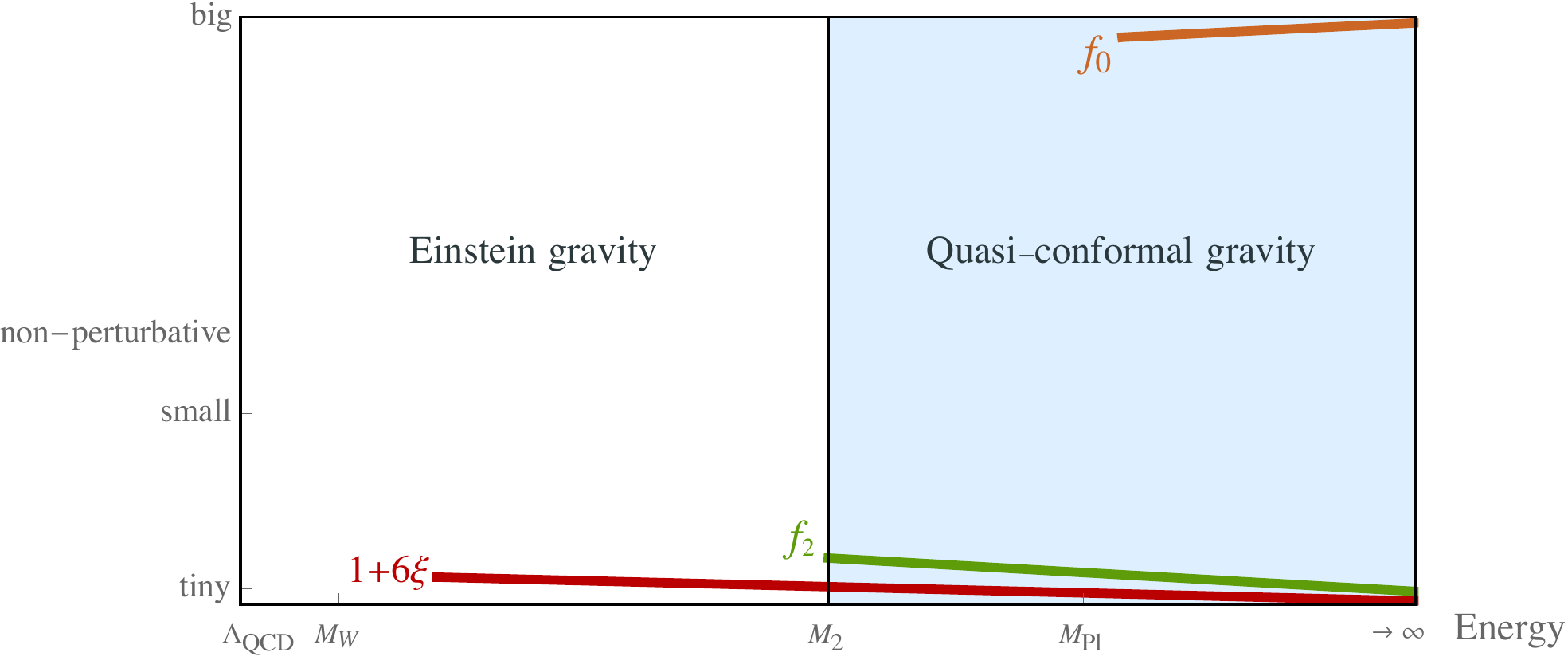} 
  \end{center}
   \caption{\em Schematic behaviour of the gravitational couplings as functions of the energy in a possible interesting scenario. 
    At high energies the theory is approximately given by conformal gravity, with small corrections (which include the UV irrelevant Einstein-Hilbert and cosmological constant terms). Both $1/f_0$ and $\delta_{ab}+6\xi_{ab}$ remain very small for the reasons given above. The coupling $f_2$ associated with the $W^2$ term is also chosen to be small both to maintain perturbativity and thus calculability and to provide interesting and potentially observable effects at the inflationary scales. The running of $f_2$ is depicted only up to the mass of the corresponding degrees of freedom, $M_2= f_2  \bp/\sqrt{2}$. A large coupling $f_0$ influences physics only at energies much above the Planck mass as  its role compared to the Einstein-Hilbert term is suppressed by $E^2/(f_0^2 \bp^2$), where $E$ is the typical energy of the process under study. Below $M_2$ the gravitational theory resembles Einstein gravity plus small corrections.
    The energy flows from the scale below which strong interactions are non-perturbative, $\Lambda_{\rm QCD}$, up to infinite energy (passing through the mass of the W-boson $M_W$, the ghost mass $M_2$ and the Planck mass $M_{\rm Pl}$).}
\label{quasi-conformal}
\end{figure}

Although flowing to conformal gravity at infinite energy can be consistent, at  finite energy conformal invariance is broken by the scale anomaly and the $R^2$ term as well as a non-vanishing value of $\delta_{ab}+6\xi_{ab}$ are generated. However, this is a multiloop effect (see \cite{Hathrell:1981zb,Hathrell:1981gz,Jack:1990eb,Salvio:2017qkx} and references therein). The full set of one-loop RGE in conformal gravity are given by 
\bea\label{eq:RGEY2}
\frac{df_2^2}{d\tau}&=&
-f_2^4\bigg(\frac{199}{15} +\frac{N_V}{5}+\frac{N_F}{20}+\frac{N_S}{60}
\bigg)\qquad 
\\
\frac{dY^a}{d\tau} &=& \frac12(Y^{\dagger b}Y^b Y^a + Y^a Y^{\dagger b}Y^b)+ 2 Y^b Y^{\dagger a} Y^b + \label{eq:RGEY22}\nonumber\\
&&+ Y^b \Tr(Y^{\dagger b} Y^a) - 3 \{ C_{2F} , Y^a\}  + \frac{15}{8}f_2^2 Y^a,~~~ \\
\frac{d\lambda_{abcd}}{d\tau} &=&  \sum_{\rm perms} \bigg[\frac18
  \lambda_{abef}\lambda_{efcd}+
  \frac38 \{\theta^A,\theta^B\}_{ab}\{\theta^A ,\theta^B\}_{cd}
  -\Tr\, Y^a Y^{\dagger b} Y^c Y^{\dagger d}+\label{eq:RGEY23}
   \nonumber
\\
&&\label{eq:RGElambda2}
+\frac5{288} f_2^4 \delta_{ab}\delta_{cd}+ \lambda_{abcd} \bigg[ \sum_{k=a,b,c,d} (Y_2^k-3  C_{2S}^k)+ 5 f_2^2\bigg]\qquad  
\eea
for $f_0\to \infty$ and $\xi_{ab}\rightarrow -\frac16 \delta_{ab}$. 
We do not show  the RGE of the gauge couplings because they are not modified by the gravitational couplings (see~\cite{Fradkin:1981iu} and~\cite{Narain:2012te,Narain:2013eea,Salvio:2014soa}).
The RGE of $f_2$ was originally derived in~\cite{Fradkin:1981iu,Shapiro-Zheksenaev,Berredo-Peixoto,Salvio:2017qkx}, while those of $Y^a$ and $\lambda_{abcd}$ were obtained in~\cite{Salvio:2017qkx}. Also, Ref.~\cite{Ohta:2015zwa} checked the RGEs of $f_2$ with functional renormalization group methods. This set of equations allows us to search for fundamental theories that enjoy total asymptotic freedom/safety: all couplings (including the gravitational ones) flow either to zero or to an interacting fixed point in the UV.

   In Fig.~\ref{quasi-conformal} a pictorial representation of a possible resulting gravitational scenario (described in the caption) is provided. 
    That behaviour suggests a new paradigm of inflation based on a quasi-conformal theory, a theory where $f_0$ is large and $\xi_{ab}\approx -\delta_{ab}/6$, which so far has been left as a very interesting future development.
 
The general RGEs in~(\ref{eq:RGEY2})-(\ref{eq:RGEY23}) can be used to address high-energy issues in the scenario presented above: e.g. the actual verification of a UV fixed point and vacuum stability.

\section{Concluding remarks}

QG, appropriately extended to include renormalizable couplings with and of a QFT, gives a renormalizable relativistic field theory of all interactions, which is predictive and computable. It has therefore attracted the interest of several researchers since decades and continues to be an important framework in the quest for a UV complete and phenomenologically viable relativistic field theory.

The price to pay is the presence of a ghost and consequently of an indefinite norm on the Hilbert space (which is implied both by renormalizability and  the requirement of having a Hamiltonian that is bounded from below). Therefore, much of this review has been dedicated  to illustrate some possible ways to address the ghost problem (such as the Dirac-Pauli quantization, the Lee-Wick approach and the possibility to introduce positively defined metrics on the Hilbert space) focusing on simple finite dimensional quantum mechanical models. The full extension of these techniques to the field theory case (and especially the QG case) has not been done yet and is an important goal for future research.

If QG is coupled to a QFT, which enjoys a UV fixed point, the whole theory can hold up to infinite energy\footnote{
If a full solution of the ghost problem in quadratic gravity is found and the theory can be made UV complete (possibly with the inclusion of matter fields) one could also have a window on strongly coupled theories through the holographic principle~\cite{tHooft:1993dmi,Susskind:1994vu} (in particular the AdS/CFT correspondence~\cite{Maldacena:1997re}) by using QG as the higher dimensional theory on an asymptotically anti-de Sitter (AdS) space. Actually, several works in this direction already appear in the literature (see e.g.~\cite{Fukuma:2001tw,Dong:2013qoa,Erdmenger:2014tba,Bhattacharjee:2015yaa}).} and might still be compatible with data. So far, potentially viable theories have only be found for $f_0^2 >0$, given that $f_0^2<0$ leads to a tachyonic instability (as is clear both in the Jordan and Einstein frame). The explicit construction of a QFT sector that satisfies all collider and cosmological bounds and explain the evidence for new physics has not been achieved yet and is an outstanding target for future research. The deep UV behaviour of the theory may be the one of a Weyl invariant theory (conformal gravity): the gravitational coupling $f_0$ and the non-minimal couplings of the scalar $\xi_{ab}$ reach the Weyl invariant values $f_0\to \infty$ and
 $\xi_{ab}\to -\delta_{ab}/6$, while all other couplings approach a UV fixed point.

\subsubsection*{Acknowledgments}
I thank   Matej Pav\u{s}i\u{c}, Ilya Shapiro, Alessandro Strumia and Hardi Veermäe for useful discussions and correspondence.
This work was supported by the ERC grant NEO-NAT.

 \vspace{1cm}
 
 \footnotesize
\begin{multicols}{2}

\end{multicols}


\begin{thebibliography}{}
\small{
\bibitem{Goroff:1985sz}
  M.~H.~Goroff and A.~Sagnotti,
  ``Quantum Gravity At Two Loops,''
  Phys.\ Lett.\  {\bf 160B} (1985) 81.
  
\bibitem{Goroff:1985th}
  M.~H.~Goroff and A.~Sagnotti,
  ``The Ultraviolet Behavior of Einstein Gravity,''
  Nucl.\ Phys.\ B {\bf 266} (1986) 709.
  
\bibitem{Utiyama:1962sn}
  R.~Utiyama and B.~S.~DeWitt,
  ``Renormalization of a classical gravitational field interacting with quantized matter fields,''
  J.\ Math.\ Phys.\  {\bf 3} (1962) 608.
  
\bibitem{Ade:2015lrj}
  P.~A.~R.~Ade {\it et al.} [Planck Collaboration],
  ``Planck 2015 results. XX. Constraints on inflation,''
  Astron.\ Astrophys.\  {\bf 594} (2016) A20 
  [\hhref{1502.02114}].
  
  
\bibitem{Starobinsky:1980te}
  A.~A.~Starobinsky,
  ``A New Type of Isotropic Cosmological Models Without Singularity,''
  Phys.\ Lett.\  {\bf 91B} (1980) 99.
  
\bibitem{Weinberg:1974tw}
 S.~Weinberg,
``Problems in Gauge Field Theories.'' In the proceedings of the XVII International Conference on High Energy Physics, editor J.~R.~Smith (Rutherford Laboratory, Chilton, Didcot, Oxfordshire), III-59.
  
\bibitem{Deser:1975nv}
  S.~Deser,
  ``The State of Quantum Gravity,''
  Conf.\ Proc.\ C {\bf 750926} (1975) 229. In the proceedings of the conference on Gauge Theories and Modern Field Theory, editors R.~Arnowitt and P.~Nath (MIT press, Cambridge, Massachusetts).
  
\bibitem{Stelle:1976gc}
  K.~S.~Stelle,
  ``Renormalization of Higher Derivative Quantum Gravity,''
  Phys.\ Rev.\ D {\bf 16} (1977) 953. 
  
    \bibitem{ostro}
M. Ostrogradsky, ``Memoires sur les \'equations diff\'erentielles relatives au probl\`eme des isop\'erim\`etres," Mem. Ac. St. Petersbourg VI (1850) 385. \href{https://babel.hathitrust.org/cgi/pt?id=mdp.39015038710128;view=1up;seq=405}{Pdf available online}.

 \bibitem{WeinbergAS} S. Weinberg, in Understanding the Fundamental Constituents of Matter, ed. A. Zichichi (Plenum 	Press, New York, 1977). S. Weinberg,  in General  Relativity: An  Einstein Centenary Survey, 	edited by S. W. Hawking and W. Israel (Cambridge University Press, 1980) pp. 790-831.

\bibitem{Biswas:2005qr}
  T.~Biswas, A.~Mazumdar and W.~Siegel,
  ``Bouncing universes in string-inspired gravity,''
  JCAP {\bf 0603} (2006) 009 
  [\hhref{hep-th/0508194}].
  
\bibitem{Stelle:1977ry}
  K.~S.~Stelle,
  ``Classical Gravity with Higher Derivatives,''
  Gen.\ Rel.\ Grav.\  {\bf 9} (1978) 353.

\bibitem{Avramidi:1985ki}
  I.~G.~Avramidi and A.~O.~Barvinsky,
  ``Asymptotic Freedom In Higher Derivative Quantum Gravity,''
  Phys.\ Lett.\  {\bf 159B} (1985) 269.
  
\bibitem{Salvio:2017qkx}
  A.~Salvio and A.~Strumia,
  ``Agravity up to infinite energy,''
  Eur.\ Phys.\ J.\ C {\bf 78} (2018) no.2,  124
  [\hhref{1705.03896}].
  
\bibitem{Buchbinder:1992rb}
  I.~L.~Buchbinder, S.~D.~Odintsov and I.~L.~Shapiro,
  ``Effective action in quantum gravity,''
  Bristol, UK: IOP (1992) 413 p
  
\bibitem{Avramidi:1986mj}
  I.~G.~Avramidi,
  ``Covariant methods for the calculation of the effective action in quantum field theory and investigation of higher derivative quantum gravity,''
    \hhref{hep-th/9510140}.
    
\bibitem{deBerredoPeixoto:2004if}
  G.~de Berredo-Peixoto and I.~L.~Shapiro,
  ``Higher derivative quantum gravity with Gauss-Bonnet term,''
  Phys.\ Rev.\ D {\bf 71} (2005) 064005
  [\hhref{hep-th/0412249}].
  
\bibitem{Shapiro:2008sf}
  I.~L.~Shapiro,
  ``Effective Action of Vacuum: Semiclassical Approach,''
  Class.\ Quant.\ Grav.\  {\bf 25} (2008) 103001 
  [\hhref{0801.0216}].
  
\bibitem{Salles:2017xsr}
  P.~Peter, F.~D.~O.~Salles and I.~L.~Shapiro,
  ``On the ghost-induced instability on de Sitter background,''
 \hhref{1801.00063}.
  
\bibitem{Kannike:2015apa}
  K.~Kannike, G.~Hütsi, L.~Pizza, A.~Racioppi, M.~Raidal, A.~Salvio and A.~Strumia,
  ``Dynamically Induced Planck Scale and Inflation,''
  JHEP {\bf 1505} (2015) 065
   [\hhref{1502.01334}].
   
\bibitem{Salvio:2017xul}
  A.~Salvio,
  ``Inflationary Perturbations in No-Scale Theories,''
  Eur.\ Phys.\ J.\ C {\bf 77} (2017) no.4,  267 
  [\hhref{1703.08012}].
  
\bibitem{Sotiriou:2008rp}
  T.~P.~Sotiriou and V.~Faraoni,
  ``$f(R)$ Theories Of Gravity,''
  Rev.\ Mod.\ Phys.\  {\bf 82} (2010) 451
  [\hhref{0805.1726}].
  
  
 
  
  
\bibitem{Holdom:2015kbf}
  B.~Holdom and J.~Ren,
  ``QCD analogy for quantum gravity,''
  Phys.\ Rev.\ D {\bf 93} (2016) no.12,  124030 
  [\hhref{1512.05305}].
  
\bibitem{Holdom:2016xfn}
  B.~Holdom and J.~Ren,
  ``Quadratic gravity: from weak to strong,''
  Int.\ J.\ Mod.\ Phys.\ D {\bf 25} (2016) no.12,  1643004
  [\hhref{1605.05006}].
  
\bibitem{Weinberg:2008zzc}
  S.~Weinberg,
  ``Cosmology,''
  Oxford, UK: Oxford Univ. Pr. (2008) 593 p.
  
\bibitem{Hindawi:1995an}
  A.~Hindawi, B.~A.~Ovrut and D.~Waldram,
  ``Consistent spin two coupling and quadratic gravitation,''
  Phys.\ Rev.\ D {\bf 53} (1996) 5583
  [\hhref{hep-th/9509142}].
  
\bibitem{Johnston:1987ue}
  D.~A.~Johnston,
  ``Sedentary Ghost Poles in Higher Derivative Gravity,''
  Nucl.\ Phys.\ B {\bf 297} (1988) 721.
  
\bibitem{Barvinsky:2017zlx}
  A.~O.~Barvinsky, D.~Blas, M.~Herrero-Valea, S.~M.~Sibiryakov and C.~F.~Steinwachs,
  ``Renormalization of gauge theories in the background-field approach,''
  \hhref{1705.03480}.
  
\bibitem{Weinberg:1995mt}
  S.~Weinberg,
  ``The Quantum theory of fields. Vol. 1: Foundations.''
  
\bibitem{Fradkin:1981iu}
  E.~S.~Fradkin and A.~A.~Tseytlin,
  ``Renormalizable asymptotically free quantum theory of gravity,''
  Nucl.\ Phys.\ B {\bf 201} (1982) 469.

  
  
\bibitem{Narain:2012te}
  G.~Narain and R.~Anishetty,
  ``Charge Renormalization due to Graviton Loops,''
  JHEP {\bf 1307} (2013) 106 
  [\hhref{211.5040}].
  
  
  
  
    
\bibitem{Narain:2013eea}
  G.~Narain and R.~Anishetty,
  ``Running Couplings in Quantum Theory of Gravity Coupled with Gauge Fields,''
  JHEP {\bf 1310} (2013) 203 
  [\hhref{1309.0473}].
  
  
  
\bibitem{Salvio:2014soa}
  A.~Salvio and A.~Strumia,
  ``Agravity,''
  JHEP {\bf 1406} (2014) 080
   [\hhref{1403.4226}].
   
   
\bibitem{Julve:1978xn}
  J.~Julve and M.~Tonin,
  ``Quantum Gravity with Higher Derivative Terms,''
  Nuovo Cim.\ B {\bf 46} (1978) 137.
  
\bibitem{Fradkin:1981hx}
  E.~S.~Fradkin and A.~A.~Tseytlin,
  ``Renormalizable Asymptotically Free Quantum Theory of Gravity,''
  Phys.\ Lett.\  {\bf 104B} (1981) 377.


\bibitem{Codello:2006in}
  A.~Codello and R.~Percacci,
  ``Fixed points of higher derivative gravity,''
  Phys.\ Rev.\ Lett.\  {\bf 97} (2006) 221301
  [\hhref{hep-th/0607128}].
  
\bibitem{Einhorn:2014bka}
  M.~B.~Einhorn and D.~R.~T.~Jones,
  ``Gauss-Bonnet coupling constant in classically scale-invariant gravity,''
  Phys.\ Rev.\ D {\bf 91} (2015) no.8,  084039
  [\hhref{1412.5572}].
  
\bibitem{Ohta:2013uca}
  N.~Ohta and R.~Percacci,
  ``Higher Derivative Gravity and Asymptotic Safety in Diverse Dimensions,''
  Class.\ Quant.\ Grav.\  {\bf 31} (2014) 015024
  [\hhref{1308.3398}].
  
     \bibitem{Pais}
A.~Pais and G.~E.~Uhlenbeck,  ``On Field theories with nonlocalized action'',  Phys.\ Rev.\  {79} (1950) 145.




  
  
\bibitem{Pagani:1987ue}
  E.~Pagani, G.~Tecchiolli and S.~Zerbini,
  ``On the Problem of Stability for Higher Order Derivatives: Lagrangian Systems,''
  Lett.\ Math.\ Phys.\  {\bf 14} (1987) 311. 
  
\bibitem{Smilga:2004cy}
  A.~V.~Smilga,
  ``Benign versus malicious ghosts in higher-derivative theories,''
  Nucl.\ Phys.\ B {\bf 706} (2005) 598 
  [\hhref{hep-th/0407231}].
  

\bibitem{Pavsic:2013noa}
  M.~Pav\u{s}i\u{c},
  ``Stable Self-Interacting Pais-Uhlenbeck Oscillator,''
  Mod.\ Phys.\ Lett.\ A {\bf 28} (2013) 1350165 
  [\hhref{1302.5257}].
  
\bibitem{Kaparulin:2014vpa}
  D.~S.~Kaparulin, S.~L.~Lyakhovich and A.~A.~Sharapov,
  ``Classical and quantum stability of higher-derivative dynamics,''
  Eur.\ Phys.\ J.\ C {\bf 74} (2014) no.10,  3072 
  [\hhref{1407.8481}].
  
\bibitem{Pavsic:2016ykq}
  M.~Pav\u{s}i\u{c},
  ``Pais-Uhlenbeck oscillator and negative energies,''
  Int.\ J.\ Geom.\ Meth.\ Mod.\ Phys.\  {\bf 13} (2016) no.09,  1630015
  [\hhref{1607.06589}].
   
\bibitem{Smilga:2017arl}
  A.~Smilga,
  ``Classical and quantum dynamics of higher-derivative systems,''
  Int.\ J.\ Mod.\ Phys.\ A {\bf 32} (2017) no.33,  1730025 
  [\hhref{1710.11538}].


\bibitem{Ivanov:2016hcm}
  M.~M.~Ivanov and A.~A.~Tokareva,
  ``Cosmology with a light ghost,''
  JCAP {\bf 1612} (2016) no.12,  018 
  [\hhref{1610.05330}].
  
\bibitem{Tokareva:2016ied}
  A.~Tokareva,
  ``Inflation with light Weyl ghost,''
  EPJ Web Conf.\  {\bf 125} (2016) 03020. 
  
\bibitem{Buchbinder:1987vp}
  I.~L.~Buchbinder and S.~L.~Lyakhovich,
  ``Canonical Quantization and Local Measure of $R^2$ Gravity,''
  Class.\ Quant.\ Grav.\  {\bf 4} (1987) 1487.
  
\bibitem{Buchbinder:1987av}
  I.~L.~Buchbinder and S.~L.~Lyakhovich,
  ``Canonical Quantization Of Theories With Higher Derivatives: Quantization Of $R^2$ Gravitation,''
  Theor.\ Math.\ Phys.\  {\bf 72} (1987) 824
   [Teor.\ Mat.\ Fiz.\  {\bf 72} (1987) 204].
   
\bibitem{Kluson:2013hza}
  J.~Kluson, M.~Oksanen and A.~Tureanu,
  ``Hamiltonian analysis of curvature-squared gravity with or without conformal invariance,''
  Phys.\ Rev.\ D {\bf 89} (2014) no.6,  064043
  [\hhref{1311.4141}].
  
\bibitem{Mannheim:2004qz}
  P.~D.~Mannheim and A.~Davidson,
  ``Dirac quantization of the Pais-Uhlenbeck fourth order oscillator,''
  Phys.\ Rev.\ A {\bf 71} (2005) 042110
  [\hhref{hep-th/0408104}].
 
\bibitem{Smilga:2005gb}
  A.~V.~Smilga,
  ``Ghost-free higher-derivative theory,''
  Phys.\ Lett.\ B {\bf 632} (2006) 433
  [\hhref{hep-th/0503213}].

\bibitem{Mannheim:2006rd}
  P.~D.~Mannheim,
  ``Solution to the ghost problem in fourth order derivative theories,''
  Found.\ Phys.\  {\bf 37} (2007) 532
  [\hhref{hep-th/0608154}].
   
\bibitem{Bolonek2006}
K.~Bolonek, P.~Kosinski, ``On Double Frequency Limit of Pais-Uhlenbeck oscillator," \hhref{quant-ph/0612009}.



\bibitem{Woodard:2006nt}
  R.~P.~Woodard,
  ``Avoiding dark energy with 1/r modifications of gravity,''
  Lect.\ Notes Phys.\  {\bf 720} (2007) 403
  [\hhref{astro-ph/0601672}].
  
\bibitem{Woodard:2015zca}
  R.~P.~Woodard,
  ``Ostrogradsky's theorem on Hamiltonian instability,''
  Scholarpedia {\bf 10} (2015) no.8,  32243 
  [\hhref{1506.02210}].



     
\bibitem{Salvio:2015gsi}
  A.~Salvio and A.~Strumia,
  ``Quantum mechanics of 4-derivative theories,''
  Eur.\ Phys.\ J.\ C {\bf 76} (2016) no.4,  227 
  [\hhref{1512.01237}].
  
  
  
  
   

  
\bibitem{Pauli}  W.~Pauli, ``On Dirac's New Method of Field Quantization", Rev.\ Mod.\ Phys.\ 15 (1943) 175.
      
\bibitem{Dirac}    P.~A.~M.~Dirac, ``The physical interpretation of quantum mechanics," 
	Proc.\ R.\ Soc.\ Lond.\ A 180, 1 (1942).
	
	
  
\bibitem{Hawking:2001yt}
  S.~W.~Hawking and T.~Hertog,
  ``Living with ghosts,"
  Phys.\ Rev.\ D {\bf 65} (2002) 103515 
  [hep-th/0107088].
  
  
\bibitem{Bender:2008gh}
  C.~M.~Bender and P.~D.~Mannheim,
  ``Exactly solvable PT-symmetric Hamiltonian having no Hermitian counterpart,''
  Phys.\ Rev.\ D {\bf 78} (2008) 025022 
  [\hhref{0804.4190}].
  
\bibitem{Bender:2007wu}
  C.~M.~Bender and P.~D.~Mannheim,
  ``No-ghost theorem for the fourth-order derivative Pais-Uhlenbeck oscillator model,''
  Phys.\ Rev.\ Lett.\  {\bf 100} (2008) 110402 
  [\hhref{0706.0207}].
  
\bibitem{Boulware:1983vw}
  D.~G.~Boulware and D.~J.~Gross,
  ``Lee-wick Indefinite Metric Quantization: A Functional Integral Approach,''
  Nucl.\ Phys.\ B {\bf 233} (1984) 1.

\bibitem{Lee:1970iw}
  T.~D.~Lee and G.~C.~Wick,
  ``Finite Theory of Quantum Electrodynamics,''
  Phys.\ Rev.\ D {\bf 2} (1970) 1033.
  
\bibitem{ColemanAcausality}
S. Coleman,  ``Acausality," in Erice 1969, Ettore Majorana School
On Subnuclear Phenomena. (New York, 1970), pp. 282?327.
  
\bibitem{Grinstein:2008bg}
  B.~Grinstein, D.~O'Connell and M.~B.~Wise,
  ``Causality as an emergent macroscopic phenomenon: The Lee-Wick O(N) model,''
  Phys.\ Rev.\ D {\bf 79} (2009) 105019 
  [arXiv:0805.2156 [hep-th]].
  
  

  
  
\bibitem{Lee:1969fy}
  T.~D.~Lee and G.~C.~Wick,
  ``Negative Metric and the Unitarity of the S Matrix,'' Nucl.\ Phys.\ B {\bf 9} (1969) 209.
  
\bibitem{Tomboulis:1980bs}
  E.~Tomboulis,
  ``Renormalizability and Asymptotic Freedom in Quantum Gravity,''
  Phys.\ Lett.\  {\bf 97B} (1980) 77.
  
\bibitem{Antoniadis:1986tu}
  I.~Antoniadis and E.~T.~Tomboulis,
  ``Gauge Invariance and Unitarity in Higher Derivative Quantum Gravity,''
  Phys.\ Rev.\ D {\bf 33} (1986) 2756.
  
\bibitem{Hasslacher:1980hd}
  B.~Hasslacher and E.~Mottola,
  ``Asymptotically Free Quantum Gravity and Black Holes,''
  Phys.\ Lett.\  {\bf 99B} (1981) 221.
  
\bibitem{Salvio:2016vxi}
  A.~Salvio,
  ``Solving the Standard Model Problems in Softened Gravity,''
  Phys.\ Rev.\ D {\bf 94} (2016) no.9,  096007 
  [\hhref{1608.01194}].
  
\bibitem{Anselmi:2017yux}
  D.~Anselmi and M.~Piva,
  ``A new formulation of Lee-Wick quantum field theory,''
  JHEP {\bf 1706} (2017) 066 
  [\hhref{1703.04584}].
  
\bibitem{Anselmi:2017lia}
  D.~Anselmi and M.~Piva,
  ``Perturbative unitarity of Lee-Wick quantum field theory,''
  Phys.\ Rev.\ D {\bf 96} (2017) no.4,  045009 
  [\hhref{1703.05563}].
  
\bibitem{Anselmi:2017ygm}
  D.~Anselmi,
  ``On the quantum field theory of the gravitational interactions,''
  JHEP {\bf 1706} (2017) 086 
  [\hhref{1704.07728}].
  
    
\bibitem{Anselmi:2018kgz}
  D.~Anselmi,
  ``Fakeons And Lee-Wick Models,''
  \hhref{1801.00915}.
  
\bibitem{Anselmi:2018ibi}
  D.~Anselmi and M.~Piva,
  ``The Ultraviolet Behavior Of Quantum Gravity,''
\hhref{1803.07777}.

\bibitem{Abe:2017abx}
  Y.~Abe, T.~Inami, K.~Izumi and T.~Kitamura,
  ``Matter scattering in $R_{\mu \nu}^2$ gravity and unitarity,''
  PTEP {\bf 2018} (2018) no.3,  031E01
  [\hhref{1712.06305}].
  
\bibitem{Donoghue:2018izj}
  J.~F.~Donoghue and G.~Menezes,
  ``Gauge Assisted Quadratic Gravity: A Framework for UV Complete Quantum Gravity,''
  \hhref{1804.04980}.

  
  
  
    
  \bibitem{Feynman}
R.P. Feynman, \href{http://cds.cern.ch/record/154856/files/pre-27827.pdf}{ ``Negative probability''}   in ``Quantum implications: Essays in honor of David Bohm", edited by B.J. Hiley and F.D. Peat (Routledge and Kegan Paul, London, 1987), Chap. 13, pp 235-248.



\bibitem{Bender:2002vv}
  C.~M.~Bender, D.~C.~Brody and H.~F.~Jones,
  ``Complex extension of quantum mechanics,''
  Phys.\ Rev.\ Lett.\  {\bf 89} (2002) 270401
   Erratum: [Phys.\ Rev.\ Lett.\  {\bf 92} (2004) 119902] 
  [\hhref{quant-ph/0208076}].
  
\bibitem{Bender:2007nj}
  C.~M.~Bender,
  ``Making sense of non-Hermitian Hamiltonians,''
  Rept.\ Prog.\ Phys.\  {\bf 70} (2007) 947 
  [\hhref{hep-th/0703096}].
  

 

 
\bibitem{Mannheim:2015hto}
  P.~D.~Mannheim,
  ``Antilinearity Rather than Hermiticity as a Guiding Principle for Quantum Theory,''
  \hhref{1512.04915}.
  
\bibitem{Raidal:2016wop}
  M.~Raidal and H.~Veermäe,
  ``On the Quantisation of Complex Higher Derivative Theories and Avoiding the Ostrogradsky Ghost,''
  Nucl.\ Phys.\ B {\bf 916} (2017) 607 
  [\hhref{1611.03498}].
  
  \bibitem{Mannheim:2017apd}
  P.~D.~Mannheim,
 ``Appropriate Inner Product for PT-Symmetric Hamiltonians,''
  Phys.\ Rev.\ D {\bf 97} (2018) no.4,  045001
  [\hhref{1708.01247}].
 
\bibitem{Strumia:2017dvt}
  A.~Strumia,
  ``Interpretation of quantum mechanics with indefinite norm,''
  \hhref{1709.04925}.
  
\bibitem{Berkin:1991nb}
  A.~L.~Berkin,
  ``Contribution of the Weyl tensor to $R^2$ inflation,''
  Phys.\ Rev.\ D {\bf 44} (1991) 1020.
  
\bibitem{Hamada:2006ei}
  K.~j.~Hamada, S.~Horata and T.~Yukawa,
  ``Space-time Evolution and CMB Anisotropies from Quantum Gravity,''
  Phys.\ Rev.\ D {\bf 74} (2006) 123502
  [\hhref{astro-ph/0607586}].
  
\bibitem{Clunan:2009er}
  T.~Clunan and M.~Sasaki,
  ``Tensor ghosts in the inflationary cosmology,''
  Class.\ Quant.\ Grav.\  {\bf 27} (2010) 165014
  [\hhref{0907.3868}].
  
  
\bibitem{Weinberg:2009wa}
  S.~Weinberg,
  ``Asymptotically Safe Inflation,''
  Phys.\ Rev.\ D {\bf 81} (2010) 083535
  [\hhref{0911.3165}].

\bibitem{Nelson:2010wp}
  W.~Nelson,
  ``Restricting Fourth Order Gravity via Cosmology,''
  Phys.\ Rev.\ D {\bf 82} (2010) 124044
  [\hhref{1012.3353}].
  
\bibitem{Deruelle:2010kf}
  N.~Deruelle, M.~Sasaki, Y.~Sendouda and A.~Youssef,
  ``Inflation with a Weyl term, or ghosts at work,''
  JCAP {\bf 1103} (2011) 040
  [\hhref{1012.5202}].
  
\bibitem{Myung:2014jha}
  Y.~S.~Myung and T.~Moon,
  ``Primordial massive gravitational waves from Einstein-Chern-Simons-Weyl gravity,''
  JCAP {\bf 1408} (2014) 061 
  [\hhref{1406.4367}].
  
\bibitem{Shapiro:2014fsa}
  I.~L.~Shapiro, A.~M.~Pelinson and F.~de O. Salles,
  ``Gravitational Waves and Perspectives for Quantum Gravity,''
  Mod.\ Phys.\ Lett.\ A {\bf 29} (2014) 1430034
  [\hhref{1410.2581}].
  
\bibitem{Myung:2015vya}
  Y.~S.~Myung and T.~Moon,
  ``Scale-invariant tensor spectrum from conformal gravity,''
  Mod.\ Phys.\ Lett.\ A {\bf 30} (2015) no.32,  1550172 
  [\hhref{1501.01749}].
  


\bibitem{Salles:2014rua}
  F.~d.~O.~Salles and I.~L.~Shapiro,
  ``Do we have unitary and (super)renormalizable quantum gravity below the Planck scale?,''
  Phys.\ Rev.\ D {\bf 89} (2014) no.8,  084054
   Erratum: [Phys.\ Rev.\ D {\bf 90} (2014) no.12,  129903] 
  [\hhref{1401.4583}].

\bibitem{Abbott:2016blz}
  B.~P.~Abbott {\it et al.} [LIGO Scientific and Virgo Collaborations],
  ``Observation of Gravitational Waves from a Binary Black Hole Merger,''
  Phys.\ Rev.\ Lett.\  {\bf 116} (2016) no.6,  061102 
  [\hhref{1602.03837}].
  

  
\bibitem{Nelson:2010ig}
  W.~Nelson,
  ``Static Solutions for 4th order gravity,''
  Phys.\ Rev.\ D {\bf 82} (2010) 104026 
  [\hhref{1010.3986}].

  
\bibitem{Lu:2015cqa}
  H.~Lu, A.~Perkins, C.~N.~Pope and K.~S.~Stelle,
  ``Black Holes in Higher-Derivative Gravity,''
  Phys.\ Rev.\ Lett.\  {\bf 114} (2015) no.17,  171601 
  [\hhref{1502.01028}].

\bibitem{Holdom:2002xy}
  B.~Holdom,
  ``On the fate of singularities and horizons in higher derivative gravity,''
  Phys.\ Rev.\ D {\bf 66} (2002) 084010 
  [\hhref{hep-th/0206219}].

  
\bibitem{Lu:2015psa}
  H.~Lü, A.~Perkins, C.~N.~Pope and K.~S.~Stelle,
  ``Spherically Symmetric Solutions in Higher-Derivative Gravity,''
  Phys.\ Rev.\ D {\bf 92} (2015) no.12,  124019 
  [\hhref{1508.00010}].
  
\bibitem{Cai:2015fia}
  Y.~F.~Cai, G.~Cheng, J.~Liu, M.~Wang and H.~Zhang,
  ``Features and stability analysis of non-Schwarzschild black hole in quadratic gravity,''
  JHEP {\bf 1601} (2016) 108 
  [\hhref{1508.04776}].
  
\bibitem{Lin:2016kip}
  K.~Lin, W.~L.~Qian, A.~B.~Pavan and E.~Abdalla,
  ``(Anti-) de Sitter Electrically Charged Black Hole Solutions in Higher-Derivative Gravity,''
  EPL {\bf 114} (2016) no.6,  60006
  [\hhref{1607.04473}].
  
\bibitem{Lu:2017kzi}
  H.~Lü, A.~Perkins, C.~N.~Pope and K.~S.~Stelle,
  ``Lichnerowicz Modes and Black Hole Families in Ricci Quadratic Gravity,''
  Phys.\ Rev.\ D {\bf 96} (2017) no.4,  046006 
  [\hhref{1704.05493}].
  
\bibitem{Goldstein:2017rxn}
  K.~Goldstein and J.~J.~Mashiyane,
  ``Ineffective Higher Derivative Black Hole Hair,''
  Phys.\ Rev.\ D {\bf 97} (2018) no.2,  024015 
  [\hhref{1703.02803}].
  
\bibitem{Kokkotas:2017zwt}
  K.~Kokkotas, R.~A.~Konoplya and A.~Zhidenko,
  ``Non-Schwarzschild black-hole metric in four dimensional higher derivative gravity: analytical approximation,''
  Phys.\ Rev.\ D {\bf 96} (2017) no.6,  064007 
  [\hhref{1705.09875}].
  
\bibitem{Stelle:2017bdu}
  K.~S.~Stelle,
  ``Abdus Salam and quadratic curvature gravity: Classical solutions,''
  Int.\ J.\ Mod.\ Phys.\ A {\bf 32} (2017) no.09,  1741012.

    
\bibitem{Holdom:2016nek}
  B.~Holdom and J.~Ren,
  ``Not quite a black hole,''
  Phys.\ Rev.\ D {\bf 95} (2017) no.8,  084034 
  [\hhref{1612.04889}].

\bibitem{Myung:2013doa}
  Y.~S.~Myung,
  ``Stability of Schwarzschild black holes in fourth-order gravity revisited,''
  Phys.\ Rev.\ D {\bf 88} (2013) no.2,  024039 
  [\hhref{1306.3725}].
  
  
    
\bibitem{Buchbinder:1989jd}
  I.~L.~Buchbinder, O.~K.~Kalashnikov, I.~L.~Shapiro, V.~B.~Vologodsky and J.~J.~Wolfengaut,
  ``The Stability of Asymptotic Freedom in Grand Unified Models Coupled to $R^{2}$ Gravity,''
  Phys.\ Lett.\ B {\bf 216} (1989) 127.
  
\bibitem{Tomboulis:2015esa}
  E.~T.~Tomboulis,
  ``Renormalization and unitarity in higher derivative and nonlocal gravity theories,''
  Mod.\ Phys.\ Lett.\ A {\bf 30} (2015) no.03n04,  1540005.
  
\bibitem{Einhorn:2015lzy}
  M.~B.~Einhorn and D.~R.~T.~Jones,
  ``Induced Gravity I: Real Scalar Field,''
  JHEP {\bf 1601} (2016) 019 
  [\hhref{1511.01481}].
  
\bibitem{Einhorn:2016mws}
  M.~B.~Einhorn and D.~R.~T.~Jones,
  ``Induced Gravity II: Grand Unification,''
  JHEP {\bf 1605} (2016) 185 
  [\hhref{1602.06290}].
  
\bibitem{Einhorn:2017icw}
  M.~B.~Einhorn and D.~R.~T.~Jones,
  ``Renormalizable, asymptotically free gravity without ghosts or tachyons,''
  Phys.\ Rev.\ D {\bf 96} (2017) no.12,  124025 
  [\hhref{1710.03795}].

\bibitem{Alvarez-Gaume:2015rwa}
  L.~Alvarez-Gaume, A.~Kehagias, C.~Kounnas, D.~Lüst and A.~Riotto,
  ``Aspects of Quadratic Gravity,''
  Fortsch.\ Phys.\  {\bf 64} (2016) no.2-3,  176
  [\hhref{1505.07657}].
  
  
\bibitem{Narain:2011gs}
  G.~Narain and R.~Anishetty,
  ``Short Distance Freedom of Quantum Gravity,''
  Phys.\ Lett.\ B {\bf 711} (2012) 128
  [\hhref{1109.3981}].
  
  
\bibitem{Narain:2016sgk}
  G.~Narain,
  ``Exorcising Ghosts in Induced Gravity,''
  Eur.\ Phys.\ J.\ C {\bf 77} (2017) no.10,  683
  [\hhref{1612.04930}].
  
\bibitem{Narain:2017tvp}
  G.~Narain,
  ``Signs and Stability in Higher-Derivative Gravity,''
  Int.\ J.\ Mod.\ Phys.\ A {\bf 33} (2018) no.04,  1850031
  [\hhref{1704.05031}].
  
  
\bibitem{Fradkin:1978yf}
  E.~S.~Fradkin and G.~A.~Vilkovisky,
  ``Conformal Invariance and Asymptotic Freedom in Quantum Gravity,''
  Phys.\ Lett.\  {\bf 77B} (1978) 262.
  
\bibitem{Zee:1983mj}
  A.~Zee,
  ``Einstein Gravity Emerging From Quantum Weyl Gravity,''
  Annals Phys.\  {\bf 151} (1983) 431.
  
\bibitem{Shapiro:1994st}
  I.~L.~Shapiro and G.~Cognola,
  ``Interaction of low-energy induced gravity with quantized matter and phase transition induced to curvature,''
  Phys.\ Rev.\ D {\bf 51} (1995) 2775
  [\hhref{hep-th/9406027}].
  
\bibitem{Hamada:2002cm}
  K.~j.~Hamada,
  ``Resummation and higher order renormalization in 4-D quantum gravity,''
  Prog.\ Theor.\ Phys.\  {\bf 108} (2002) 399 
  [\hhref{hep-th/0203250}].
  
\bibitem{Hamada:2009hb}
  K.~j.~Hamada,
  ``Renormalizable 4D Quantum Gravity as A Perturbed Theory from CFT,''
  Found.\ Phys.\  {\bf 39} (2009) 1356 
  [\hhref{0907.3969}].
  
  
\bibitem{Donoghue:2016xnh}
  J.~F.~Donoghue,
  ``Conformal model of gravitons,''
  Phys.\ Rev.\ D {\bf 96} (2017) no.4,  044006 
  [\hhref{1609.03524}].
  
\bibitem{Alvarez:2017spt}
  E.~Alvarez, J.~Anero and S.~Gonzalez-Martin,
  ``Quadratic gravity in first order formalism,''
  JCAP {\bf 1710} (2017) no.10,  008
  [\hhref{1703.07993}].
  
\bibitem{Giudice:2014tma}
  G.~F.~Giudice, G.~Isidori, A.~Salvio and A.~Strumia,
  ``Softened Gravity and the Extension of the Standard Model up to Infinite Energy,''
  JHEP {\bf 1502} (2015) 137
  [\hhref{1412.2769}].
  
\bibitem{Holdom:2014hla}
  B.~Holdom, J.~Ren and C.~Zhang,
  ``Stable Asymptotically Free Extensions (SAFEs) of the Standard Model,''
  JHEP {\bf 1503} (2015) 028
  [\hhref{1412.5540}].
  
\bibitem{Pelaggi:2015kna}
  G.~M.~Pelaggi, A.~Strumia and S.~Vignali,
  ``Totally asymptotically free trinification,''
  JHEP {\bf 1508} (2015) 130
  [\hhref{1507.06848}].
  
\bibitem{Mann:2017wzh}
  R.~Mann, J.~Meffe, F.~Sannino, T.~Steele, Z.~W.~Wang and C.~Zhang,
  ``Asymptotically Safe Standard Model via Vectorlike Fermions,''
  Phys.\ Rev.\ Lett.\  {\bf 119} (2017) no.26,  261802
  [\hhref{1707.02942}].
  
\bibitem{Pelaggi:2017abg}
  G.~M.~Pelaggi, A.~D.~Plascencia, A.~Salvio, F.~Sannino, J.~Smirnov and A.~Strumia,
  ``Asymptotically Safe Standard Model Extensions?,''
  \hhref{1708.00437}.
  
\bibitem{Hathrell:1981zb}
  S.~J.~Hathrell,
  ``Trace Anomalies and $\lambda \phi^4$ Theory in Curved Space,''
  Annals Phys.\  {\bf 139} (1982) 136. 
  
\bibitem{Hathrell:1981gz}
  S.~J.~Hathrell,
  ``Trace Anomalies and {QED} in Curved Space,''
  Annals Phys.\  {\bf 142} (1982) 34. 
  
\bibitem{Jack:1990eb}
  I.~Jack and H.~Osborn,
  ``Analogs for the $c$ Theorem for Four-dimensional Renormalizable Field Theories,''
  Nucl.\ Phys.\ B {\bf 343} (1990) 647.
  

  

\bibitem{Shapiro-Zheksenaev}
I. L. Shapiro and A. G. Zheksenaev, ``Gauge dependence in higher derivative quantum gravity and the conformal anomaly problem", Phys. Lett. B 324 (1994) 286. 

\bibitem{Berredo-Peixoto}
G. de Berredo-Peixoto, I.L. Shapiro, ``Conformal quantum gravity with the Gauss-Bonnet term", Phys. Rev. D70 (2003) 044024 [\hhref{hep-th/0307030}]

\bibitem{Ohta:2015zwa}
  N.~Ohta and R.~Percacci,
  ``Ultraviolet Fixed Points in Conformal Gravity and General Quadratic Theories,''
  Class.\ Quant.\ Grav.\  {\bf 33} (2016) 035001
  [\hhref{1506.05526}].

\bibitem{tHooft:1993dmi}
  G.~'t Hooft,
  ``Dimensional reduction in quantum gravity,''
  Conf.\ Proc.\ C {\bf 930308} (1993) 284
  [\hhref{gr-qc/9310026}].

\bibitem{Susskind:1994vu}
  L.~Susskind,
  ``The World as a hologram,''
  J.\ Math.\ Phys.\  {\bf 36} (1995) 6377 
  [\hhref{hep-th/9409089}].
  
\bibitem{Maldacena:1997re}
  J.~M.~Maldacena,
  ``The Large N limit of superconformal field theories and supergravity,''
  Int.\ J.\ Theor.\ Phys.\  {\bf 38} (1999) 1113
   [Adv.\ Theor.\ Math.\ Phys.\  {\bf 2} (1998) 231]
  [\hhref{hep-th/9711200}].
  
 
\bibitem{Fukuma:2001tw}
  M.~Fukuma and S.~Matsuura,
  ``Holographic renormalization group structure in higher derivative gravity,''
  Prog.\ Theor.\ Phys.\  {\bf 107} (2002) 1085
  [\hhref{hep-th/0112037}].
  
        

\bibitem{Dong:2013qoa}
  X.~Dong,
  ``Holographic Entanglement Entropy for General Higher Derivative Gravity,''
  JHEP {\bf 1401} (2014) 044
  [\hhref{1310.5713}].
 
        
\bibitem{Erdmenger:2014tba}
  J.~Erdmenger, M.~Flory and C.~Sleight,
  ``Conditions on holographic entangling surfaces in higher curvature gravity,''
  JHEP {\bf 1406} (2014) 104
  [\hhref{1401.5075}].
  
\bibitem{Bhattacharjee:2015yaa}
  S.~Bhattacharjee, S.~Sarkar and A.~C.~Wall,
  ``Holographic entropy increases in quadratic curvature gravity,''
  Phys.\ Rev.\ D {\bf 92} (2015) no.6,  064006
  [\hhref{1504.04706}].
  
  
  
  
  
  
  
  

   
  


}
  
  
 
  
  
 







\end{thebibliography}
\end{document}